\documentclass[11pt,twoside,a4paper]{report}

\usepackage{hyperref}
\usepackage{amsmath}
\usepackage{amssymb}
\usepackage{xspace}
\usepackage[english]{babel}
\usepackage{braket}
\usepackage{cite}

\usepackage{graphicx}
\usepackage{subfig}

\usepackage{float}

\begin{document}
	
\title{Notes on the Kitaev $\Leftrightarrow$ Tight-binding correspondence\\
\large extracted from a BSc project in Physics presented at Universidade de Aveiro}
\date{July 14, 2017}
\author{Rui Carlos Andrade Martins\\ Departamento de F\'isica, Universidade de Aveiro}
\maketitle

\begin{abstract}
	In this project, we study the properties of a non-trivial topological system which exhibits localized edge states. In our study, we address the Kitaev chain, a one-dimensional chain of atoms deposited on top of a p-wave superconductor that induces superconductivity in the chain by proximity effect. We establish a correspondence between the Kitaev chain and a tight-binding lattice with a particular geometry for a particular case of the system parameters. This correspondence allows one to find the exact energy levels and eigenstates of the $t=|\Delta|$ Kitaev chain when $\mu=0$ for an arbitrary chain size.
\end{abstract}

\pagenumbering{roman}
\tableofcontents

\cleardoublepage
\listoffigures

\cleardoublepage
\pagenumbering{arabic}

\chapter{Introduction}
In 1997, Alexei Y. Kitaev proposed the concept of topological quantum computation \cite{kitaev_1997}. This theoretical model employs the properties of \textit{non-abelian} anyons to construct the logic gates necessary for the computer to function. It also provides a way to avoid decoherence at the physical level since the encoding of information using a topological quantum computer is done in a non-local fashion. In mathematics, very briefly, the concept of topological equivalence classifies surfaces in $\mathbb{R}^n$ by the number of holes in them \cite{decarlo1996topological}. Two objects are topologically equivalent if one can be transformed into the other by a homeomorphism \cite{armstrong2013basic}. For example, a spherical surface is topologically equivalent to the surface of an ellipsoid since one shape can be deformed into the other continuously. Topological invariants are quantities or objects preserved during an homeomorphism. In physics, there exist quantities which are directly connected to the topology of the system, and thus only change if it undergoes a transition between topologically different phases, meaning that these quantities remain unaltered by adiabatic changes in the parameters. In the systems we are going to study, a smooth deformation of the physical parameters which leave the band gap finite in the band structure do not change the topological phase of the system. In other words, the topological phase shifts whenever the gap closes. Knowing this, we can identify topological transitions by looking at the band structure of such systems. Certain topologically invariant quantities, such as the Zak's phase \cite{delplace2011zak}, allow to distinguish between topologically trivial and non-trivial phases of the system. A physical signature of non-trivial topological phases is given by the presence of localized states at the edges of a system with open-boundaries, which are topologically protected against perturbations. The Majorana states that we will study in this work are an example of such topological states.

In this project we will study one of the simplest model in which Majorana fermions, which are quasiparticles that obey \textit{non-abelian} statistics, appear. This system consists of a one-dimensional tight-binding chain deposited on top of a effectively spinless p-wave superconductor, and was introduced originally by Alexei Y. Kitaev \cite{kitaev2001unpaired}. As shown by Fu and Kane \cite{fu2008superconducting} such a system might occur for the surface states of a topological insulator when brought into  tunneling contact with an ordinary s-wave superconductor . In order to lay the foundations for our analysis, we will first introduce the tight-binding approximation and study simple 1D tight-biding systems in Chapter 2, finding their electronic band structure. We will consider a tight-binding Hamiltonian for a chain of atoms with open boundary conditions and for a quantum ring, which is a chain of atoms with periodic boundary conditions, both with and without magnetic flux. In Chapter 3 we will introduce the Kitaev chain and study some of its properties, such as its topological characterization and the existence of Majorana end states. Still in the same chapter we establish a correspondence between the Kitaev chain system and a specific tight-binding problem, for a special case of the parameters. We will study this correspondence for the particular case of the three site Kitaev chain, providing the groundwork required to establish the correspondence for a chain with a larger number of sites. In the last section of this chapter, we provide a set of rules that allows one to construct the tight-binding lattices and relate the solutions of both problems. Finally, we present the conclusions on Chapter 4.
\chapter{The Tight-Binding Approximation}
In this Chapter we will introduce the tight-binding approximation and study 1D tight-binding models with both open and periodic boundary conditions and also in the presence of flux in the latter case. This Chapter is based on references \cite{Sigrist} and \cite{kittel2005introduction}.
\section{The Method}
The tight-binding model is an approach to the calculation of the electronic band structure of a lattice that relies on the fact that electrons are either tightly bound to the atoms or that these are well separated. Very briefly, the method consists of expressing the eigenstates of a lattice in terms of the localized atomic orbitals of the atoms.

Let us start by defining localized atomic orbitals $\phi_n(\vec{r}-\vec{R}_j)$, where $\vec{R}_j$ is the lattice vector correspondent to site $j$, $\vec{r}$ is the position vector. These are the eigenstates of the atomic Hamiltonian $H_{a}(\vec{R}_j)$, which is the Hamiltonian of a non-interacting atom located at site $j$. We have
\begin{equation}
H_{a}(\vec{R}_j)\phi_n(\vec{r}-\vec{R}_j)=\epsilon_n\phi_n(\vec{r}-\vec{R}_j)	,
\end{equation}
where $\epsilon_n$ is the energy and $n$ stands for the set of all necessary quantum numbers to fully describe an orbital of the atom. In the context of our work, we will only consider one type of  atomic orbital to be important to model the transport properties of the system, so we will drop the index $n$. Now, we assume that these atomic orbitals quickly decay so that they overlap only slightly, i.e, they are almost orthogonal, and so, they constitute a complete set of orthogonal functions. One can, therefore, describe these functions in terms of the Wannier functions, $\Phi(\vec{R}_j)$, of the lattice .

Bloch's theorem states the eigenfunctions of a system with a periodic repeating potential are Bloch's functions, which are plane waves with the same periodicity as the lattice. These functions can be expanded on the basis of the Wannier functions as follows

\begin{equation}
\ket{\psi_{\vec k}}=\sum_{j}e^{i\vec{k}\cdot \vec{R_j}}\ket{\Phi(\vec{R_j})}
\end{equation}
where $\vec{k}$ is a wave momentum and a vector from the reciprocal lattice contained in the first Brillouin zone. Eigenfunctions corresponding to different $\vec{k}$ vectors are orthogonal. As we said before, we can approximate the atomic orbitals by this set of Wannier functions (such that the atomic orbital of site $j$  will be approximated by the Wannier function corresponding to site $j$) since they constitute a complete set of orthogonal functions,

\begin{equation}
\ket{\psi_{\vec k}}=\sum_{j}e^{i\vec{k}\cdot\vec{R_j}}\ket{\phi(\vec{r}-\vec{R}_j)}.
\end{equation}
In the context of our work we will only focus on one-dimensional systems. With this in mind, let us write  the solution of the one-dimensional problem in second quantization formalism using the result above,

\begin{equation}
	c_k^\dagger\ket{0}=\sum_j e^{ikj}c_j^\dagger\ket{0}	
\end{equation}
where $c_k^\dagger$ is the creation operator of an extended electronic state, a Bloch function, with a well defined momentum $k$, $c_j^\dagger$ is the creation operator of a localized electron at site $j$, a Wannier state, and $\ket{0}$ is the vacuum state of the system. Notice that we dropped the vector notation for $k$ given that we are considering 1D systems. Also, we assumed the lattice parameter $a$ to be $1$, so that $\vec{k}\cdot\vec{R_j}=kja=kj$.

Having stated all this, we are ready to write the tight-binding Hamiltonian of a lattice in second quantization. One can write the lattice Hamiltonian as

\begin{equation}
	H=\sum_{R_j} H_W (R_j)+\Delta{H},
\end{equation}
where $H_W (R_j)$ is a Hamiltonian whose eigenfunctions are Wannier functions centered on $R_j$ and $\Delta{H}$ is a correction term associated with the overlap between atomic orbitals, necessary to reproduce the full Hamiltonian. As we said before, we assume that only one orbital of the atom is present, and therefore, only one Wannier state is important to describe the transport properties of the system. With this in mind, we define

\begin{eqnarray}
	-t_{i,j}&=&\bra{0}c_i\Delta{H}c_j^\dagger\ket{0}=\int dr \phi(\vec{r}-\vec{R}_i)^*\Delta{H}\phi(\vec{r}-\vec{R}_j),
	\nonumber
	\\
	\epsilon_{j}&=&\bra{0}c_j\sum_{j^{'}} H_W (R_{j^{'}})c_j^\dagger\ket{0}
	=\int dr \phi(\vec{r}-\vec{R}_j)^*\sum_{j^{'}} H_W (R_{j^{'}})\phi(\vec{r}-\vec{R}_j)
	\nonumber
	\\
	&=&\int dr \phi(\vec{r}-\vec{R}_j)^*\delta_{j j^{'}}H_W (R_{j^{'}})\phi(\vec{r}-\vec{R}_j),
\end{eqnarray}
where we took into account that the atomic orbital of site $j$ was approximated by the  the Wannier state of site $j$, so that $\phi(\vec{r}-\vec{R}_j)\rightarrow c_j^\dagger\ket{0}$. Now we can write the Hamiltonian in second quantization formalism as follows,

\begin{equation}
	H=\sum_{i,j}(\delta_{ij}\epsilon_{j}-t_{i,j})c_i^\dagger c_j,
\end{equation}
where $\epsilon_{j}$ is the energy associated with an electron occupying the orbital of the atom  at site $j$ and $t_{i,j}$ is the hopping parameter, which in turn is the kinetic energy parameter associated with the probability of transition of an electron from the atom at site $j$ to the atom at site $i$.

Finally, we will assume the system to be isotropic, as we have been implying already, since the atomic orbital functions are similar and only differ in where they are centered, and we have
$\epsilon_j=\epsilon$. Moreover, we only consider hopping between nearest-neighboring sites, so we have

\begin{equation}
\left\{\begin{matrix}
t_{i,j}=t, |i-j|=1,\\
t_{i,j}=0, \text{otherwise}.
\end{matrix}\right.
\end{equation} 
This way, the Hamiltonian becomes

\begin{equation}
	H=\epsilon\sum_{j}c_j^\dagger c_j-t\sum_{j}(c_j^\dagger c_{j+1}+c^\dagger_{j+1}c_j).
\end{equation}
 
 \section{Quantum ring - chain with periodic boundary conditions}
 Now we focus on studying a quantum ring, which is a structure composed by $N$ atoms in a periodic arrangement.
 
  \begin{figure}[h]
 	\centering
 	\includegraphics[scale=0.6]{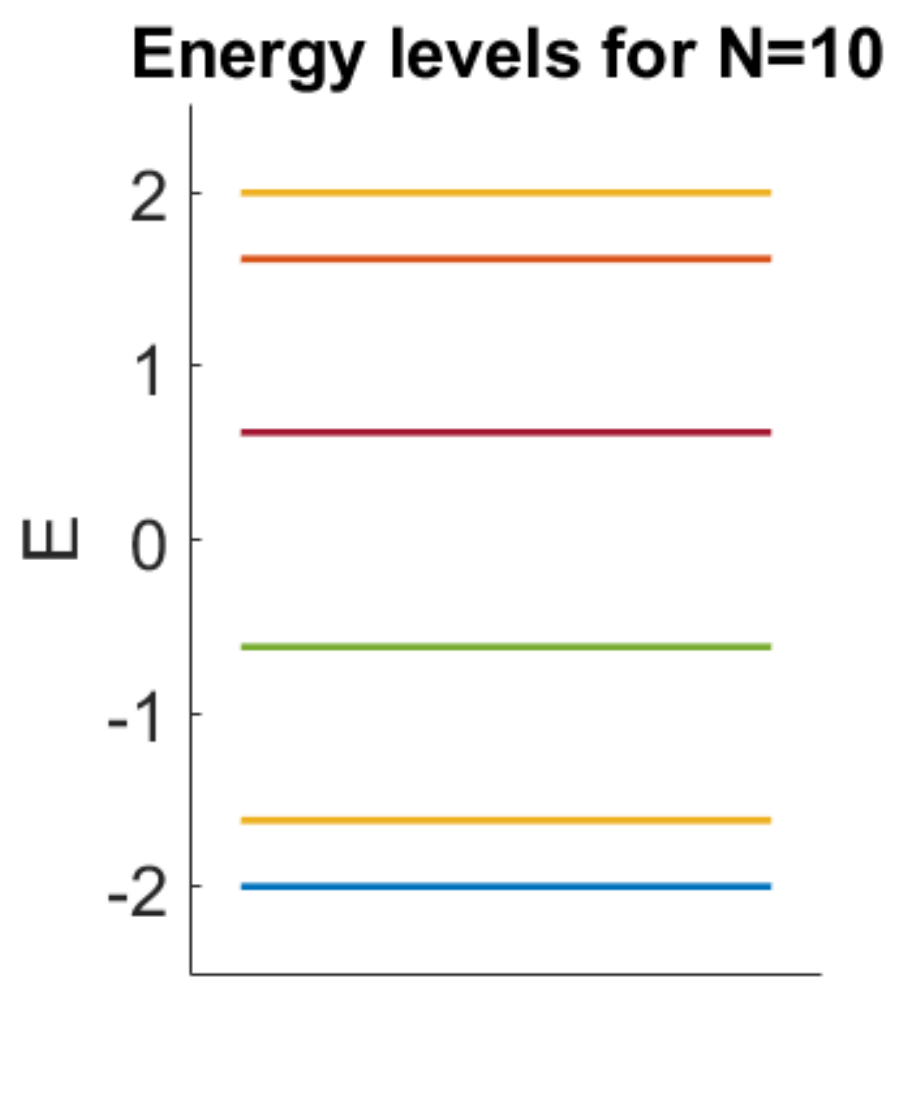}
 	\caption{Energy levels for an quantum ring with $N=10$. Energy is in units of $t$.}
 	\label{Energy_closed} 
 \end{figure}
 
 \subsection{Quantum ring without flux}
 
 \begin{figure}[!tbp]
 	\subfloat[ ]{\includegraphics[width=0.45\textwidth]{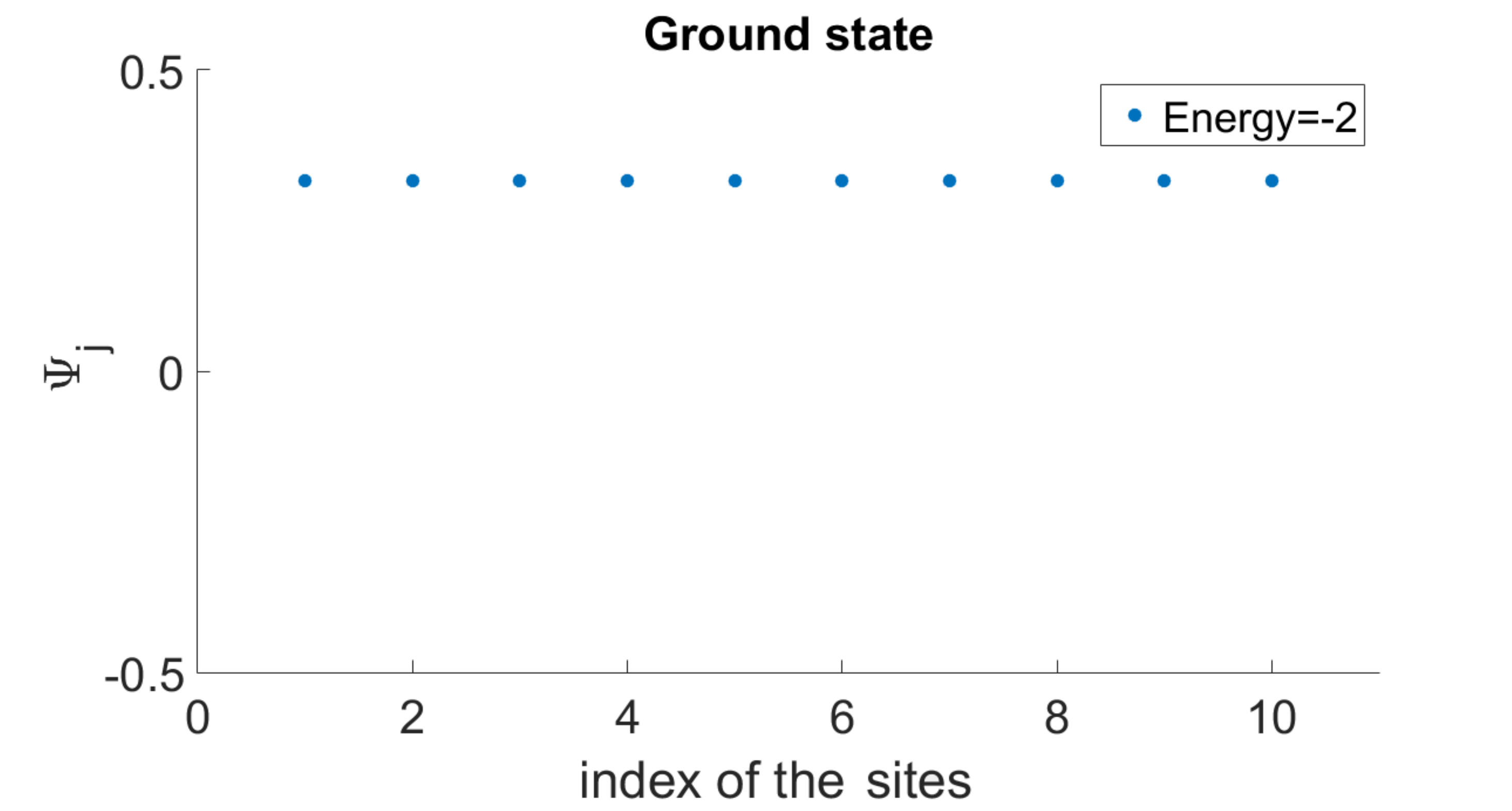}\label{b}}
 	\qquad
 	\subfloat[ ]{\includegraphics[width=0.45\textwidth]{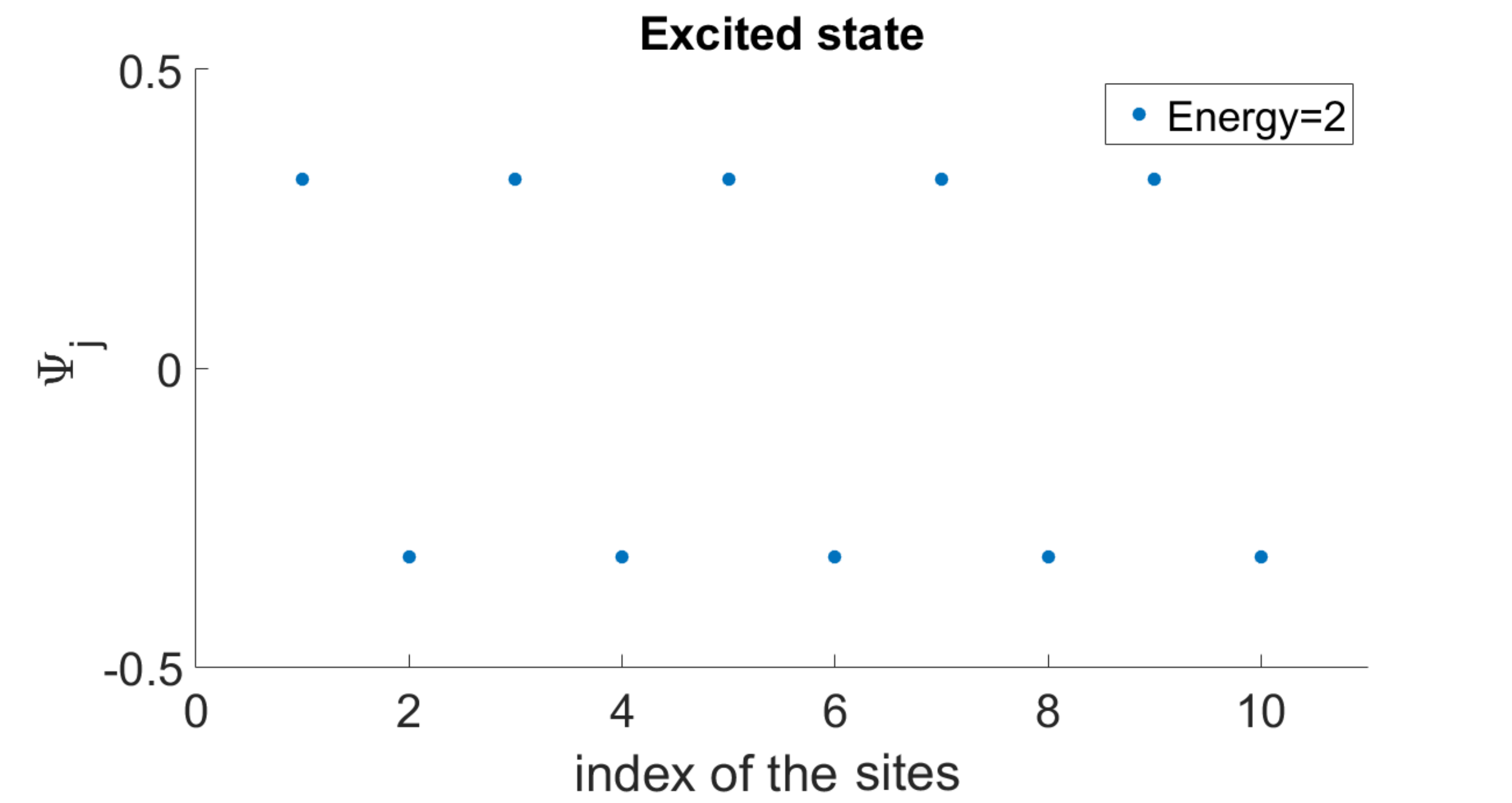}\label{c}}
 	\hfill
 	\caption{Plot of the amplitude distribution of (a) the ground eigenstate with energy $-2t$ and (b) highest energy eigenstate with energy $2t$ for a quantum ring with ten sites and $t=1$.}
 \end{figure}
 We consider a system described by the following Hamiltonian
 
 \begin{equation}
 	H=-t\sum_{j=1}^N(c_j^\dagger c_{j+1}+c^\dagger_{j+1}c_j),
 \end{equation}
 where we assume periodic boundary conditions, such that site $1$ is equivalent to site $N+1$, and that the on-site energy $\epsilon$ is $0$. Since the Hamiltonian has translational invariance it is diagonalized by Bloch's states, therefore we perform a Fourier transform to go into k-space,
 
 \begin{equation}
	c_j^\dagger=\dfrac{1}{\sqrt{N}}\sum_{k}e^{-ikj}c_k^\dagger
	,\qquad
	c_j=\dfrac{1}{\sqrt{N}}\sum_{k}e^{ikj}c_k,
	\label{Bloch_j}
 \end{equation} 
 where, due to periodic boundary conditions, $k=\frac{2\pi n}{N}$, which results from the equality 

\begin{equation}
		c_1^\dagger=\dfrac{1}{\sqrt{N}}\sum_{k}e^{-ik}c_k^\dagger=\dfrac{1}{\sqrt{N}}\sum_{k}e^{-ik(N+1)}c_k^\dagger=c_{N+1}^\dagger,
\end{equation}
where

\begin{equation}
\left\{\begin{matrix}
n=-\frac{N}{2},-\frac{N}{2}+1,...,\frac{N}{2}-1,\text{when N is even}\\
n=-\frac{N-1}{2},-\frac{N-1}{2}+1,...,\frac{N-1}{2},\text{when N is odd}
\end{matrix}\right.
\label{eq_fixe}
\end{equation}
so that $k$ is contained in the first Brillouin zone $]\-\frac{\pi}{a},\frac{\pi}{a}]$. By using the orthogonality relation

\begin{equation}
	\sum_{j}e^{i(k-k^{'})j}=\delta_{kk^{'}}N
	\label{eq_orto},
\end{equation}
we arrive at the following diagonalized Hamiltonian

\begin{equation}
	H=-2t\sum_{k}\cos(k)c_k^\dagger c_k,
\end{equation}
and the band structure of the quantum ring. In Fig. \ref{Energy_closed}, we plot the energy levels of a quantum ring with ten sites and in Fig. \ref{b} and \ref{c}, we plot the amplitude distribution of the ground and highest energy eigenstates respectively.

\subsection{Quantum ring with flux}
 
  \begin{figure}
 	\centering
 	\includegraphics[scale=0.45]{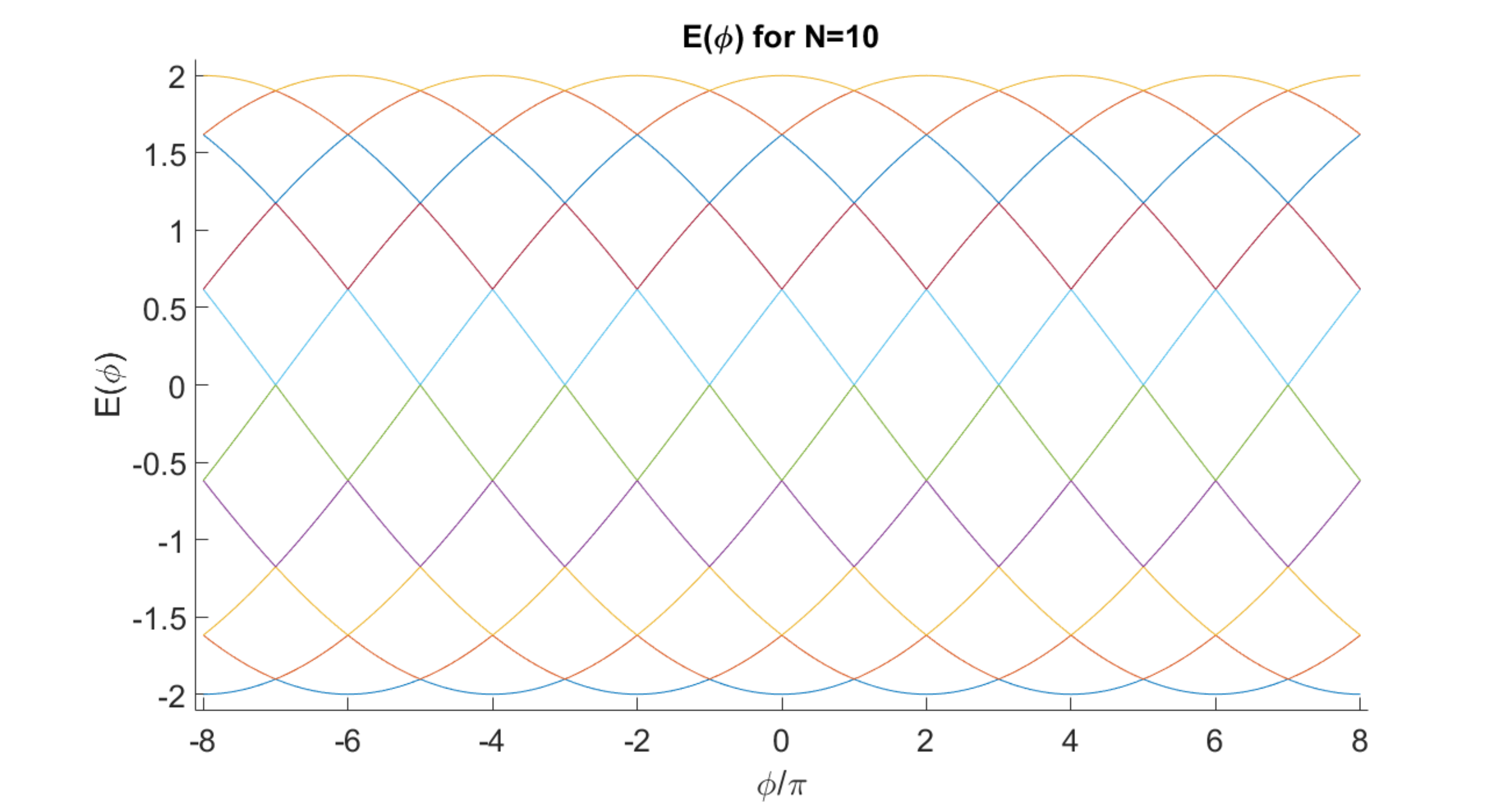}
 	\caption{Energy dispersion of a ten site quantum ring as a function of the reduced magnetic flux $\phi$. The energy is in units of $t$.}
 	\label{flux}
 \end{figure}

 Now we examine the ring while it is under a perpendicular magnetic field. Let us consider the magnetic field pierces the ring in a region where the amplitude of the wavefunction is almost null, so no forces are exerted on the electrons. Even so, due to the Aharonov-Bohm effect \cite{aharonov1961further}, the wave function of the electron when the ring is under a magnetic field is altered by a phase compared to the wavefunction of the ring in the absence of a magnetic field. The phase shift of the wavefunction when the electron travels along a path $L$ is called the Peierls phase, and it is given by
 
 \begin{equation}
 exp\left[ i\frac{q}{\hbar}\int_{L}\vec{A}\cdot d\vec{l}\right].
 \end{equation}
 where $q$ is the electron charge, $\vec{A}$ is the vector potential and $\hbar$ the reduced Planck's constant. We can relate the Wannier state of an electron when the ring is under a magnetic field, $\ket{j}_B$, to the Wannier state in the absence of field $\ket{j}_0$ by
  
\begin{equation}
	\ket{j}_B=exp\left[ i\frac{q}{\hbar}\int_{j-1}^j\vec{A}\cdot d\vec{l}\right]\ket{j}_0
\end{equation}
where the path of integration will be the straight line connecting nearest-neighbors. Since the system must be gauge invariant, we consider the gauge where the Peierls phase is the same on every path $L$ between nearest-neighbors, in order to preserve the translational symmetry, so that

\begin{equation}
	\int_{j-1}^{j}\vec{A}\cdot d\vec{l}=\frac{1}{N}\int_{1}^N\vec{A}\cdot d\vec{l}=\frac{\phi}{N},
\end{equation}
where $\phi$ is the total magnetic flux piercing the ring. Taking into account this choice of gauge, the Hamiltonian becomes

\begin{equation}
 H=-t\sum_{j=1}^N e^{i\frac{\phi^{'}}{N}}c_j^\dagger c_{j+1}+e^{-i\frac{\phi^{'}}{N}}c^\dagger_{j+1}c_j,
 \end{equation}
 where $\phi^{'}=2\pi\frac{\phi}{\phi_0}$ is the reduced flux and $\phi_0=\frac{h}{q}$ is the flux quantum. Considering this choice of gauge, the Hamiltonian has translational invariance and so Bloch waves are its eigenstates. The method to find the solution follows the same procedure of the case without flux, and we get
 
 \begin{equation}
 H=-2t\sum_{k}cos(k+\frac{\phi^{'}}{N})c_k^\dagger c_k.
 \end{equation}
 
 In Fig. \ref{flux}, we plot the energy as a function of the reduced magnetic flux $\phi^{'}$, for an open chain with $N=10$ and hopping parameter $t=-1$.
 
 \section{Chain with open boundary conditions}
 Now we study an $N$-site tight-binding chain with open boundaries, meaning that the amplitude of the wavefunction must be zero beyond the ends of the chain. We consider the following Hamiltonian
 
 \begin{equation}
 H=-t\sum_{j=1}^{N-1}(c_j^\dagger c_{j+1}+c^\dagger_{j+1}c_j).
 \end{equation}
 The wavefunctions in the bulk of the chain should be similar to the case of the quantum ring (at least, if we consider a big enough open chain). However, due to the boundary conditions, only the wavefunctions that obey the following conditions are allowed
 
 \begin{equation}
 \left\{\begin{matrix}
 E_{k^{'}}\psi_1=-t\psi_2\\
 E_{k^{'}}\psi_N=-t\psi_{N-1}
 \end{matrix}\right.
 \end{equation}
 where $\psi_j$ is the amplitude of the eigenfunction at site $j$ and $E_{k^{'}}$ its energy. To solve this problem, we add two virtual sites to the chain, site $j=0$ and site $j=N+1$ and impose that the amplitude of the wavefunction at these sites must be zero. We assume that the eigenfunctions are linear combinations of Bloch waves of the same energy and opposite momenta of the form
 
\begin{equation}
\ket{k,open}=\frac{1}{\sqrt{2}}(\ket{k}-\ket{-k})=C\left(\sum_{j}\sin(kj)c^\dagger_j\right)\ket{\emptyset}
\end{equation}
where $\ket{k,open}$ is the eigenstate of the system, $\ket{k}$ is a Bloch wave with momentum $k$, $\ket{\emptyset}$ is the vacuum and $C$ is some normalization constant. These linear combinations originate standing waves that allow both conditions to be satisfied. On the one hand, they allow the amplitude of the wavefunctions to vanish at the border sites and, on the other hand, they are still eigenstates of the bulk Hamiltonian, since they combine Bloch waves with the same energy.

\begin{figure}
	\centering
	\includegraphics[scale=0.6]{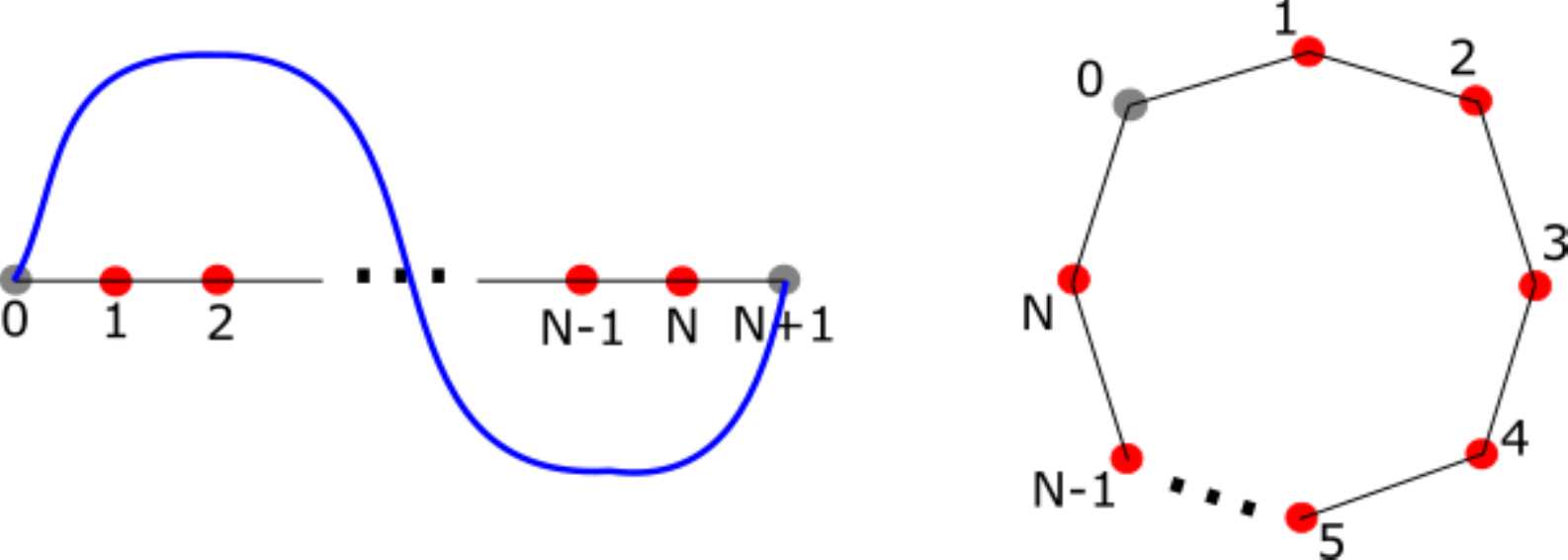}
	\caption{Pictorial representation of the lowest energy odd parity standing wave and equivalent tight-binding $N+1$ ring at the right.}
	\label{x}
\end{figure}

\begin{figure}
	\centering
	\begin{minipage}{.5\textwidth}
		\centering
		\includegraphics[width=.8\linewidth]{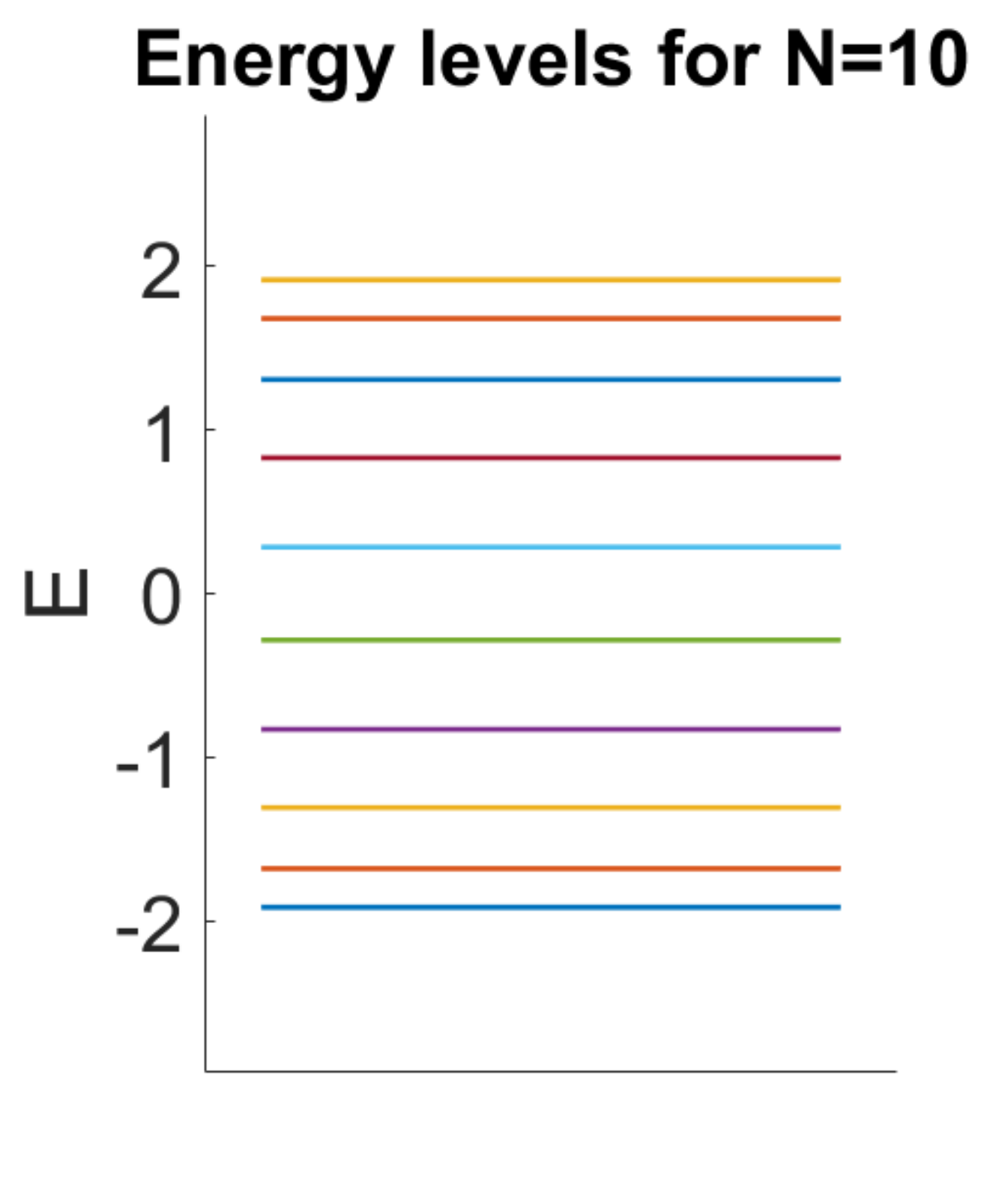}
		\caption{Energy levels of a tight-binding open chain with ten sites. The energy is in units of $t$.}
		\label{energy_open}
	\end{minipage}%
	\begin{minipage}{.5\textwidth}
		\centering
		\includegraphics[width=1.0\linewidth]{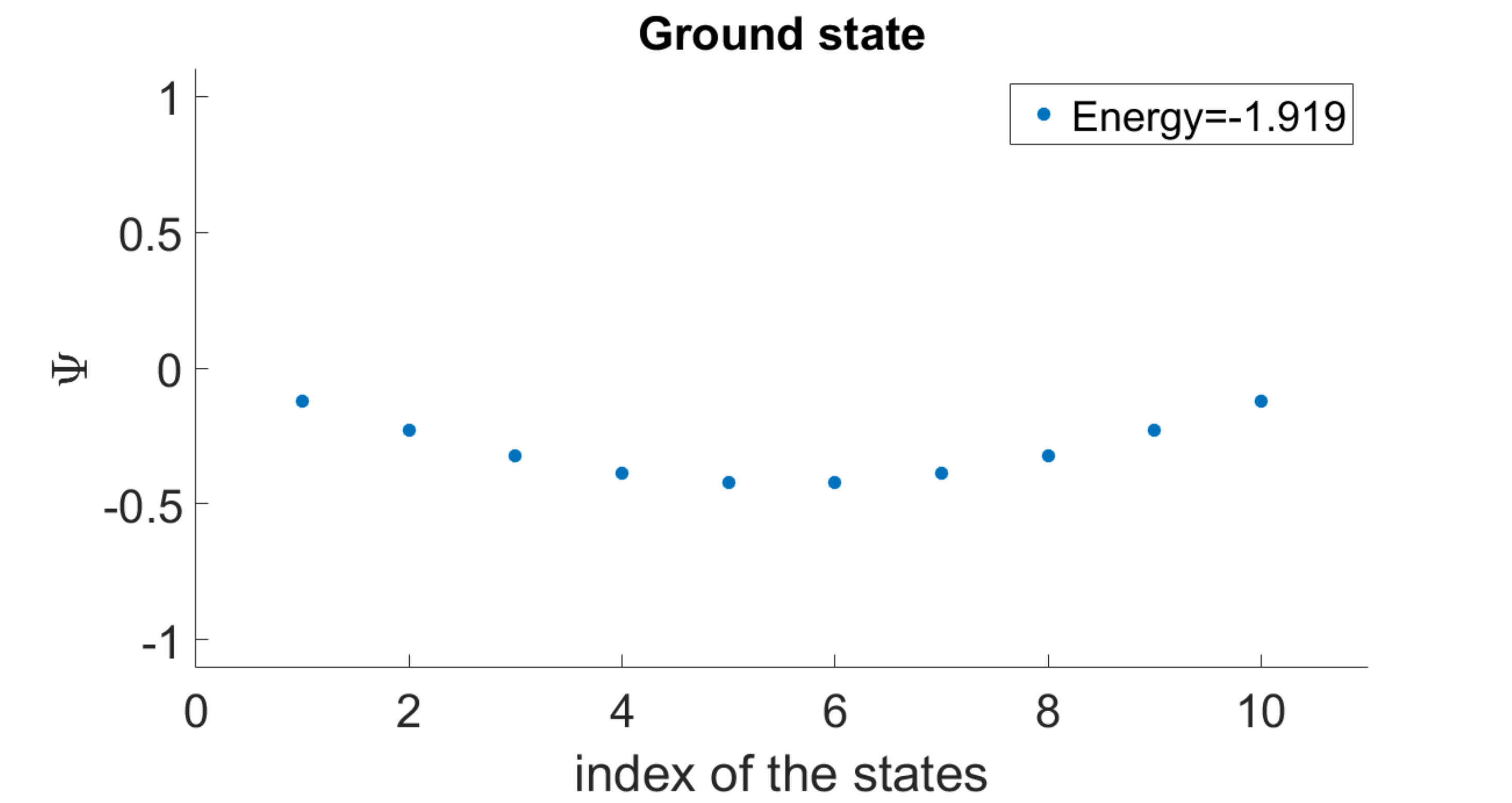}
		\caption{Amplitude distribution for the ground state of Fig. \ref{energy_open}.}
		\label{ground_open}
	\end{minipage}
\end{figure}

The only thing left to find the solution is to determine the normalization constant $C$, that results from the fact that only certain $k$ are allowed as a solution. We should distinguish between the case where the solutions are standing waves with odd parity and with even parity with relation to the center of the chain. Let us start by taking into account the odd parity solutions (see Fig. \ref{x}). In this case, the problem corresponds to one of a tight-binding ring with $N+1$ sites, with boundary conditions $\ket{0}=\ket{N+1}$ and the added condition that the amplitude of the wavefunction is zero at site $j=0$. We obtain the quantization for $\ket{k,open}$ by applying the boundary condition
\begin{equation}
	\sin(k(N+1))=2\pi m
	,\qquad
	m=0,...,N
\end{equation}
\begin{figure}
	\centering
	\includegraphics[scale=0.6]{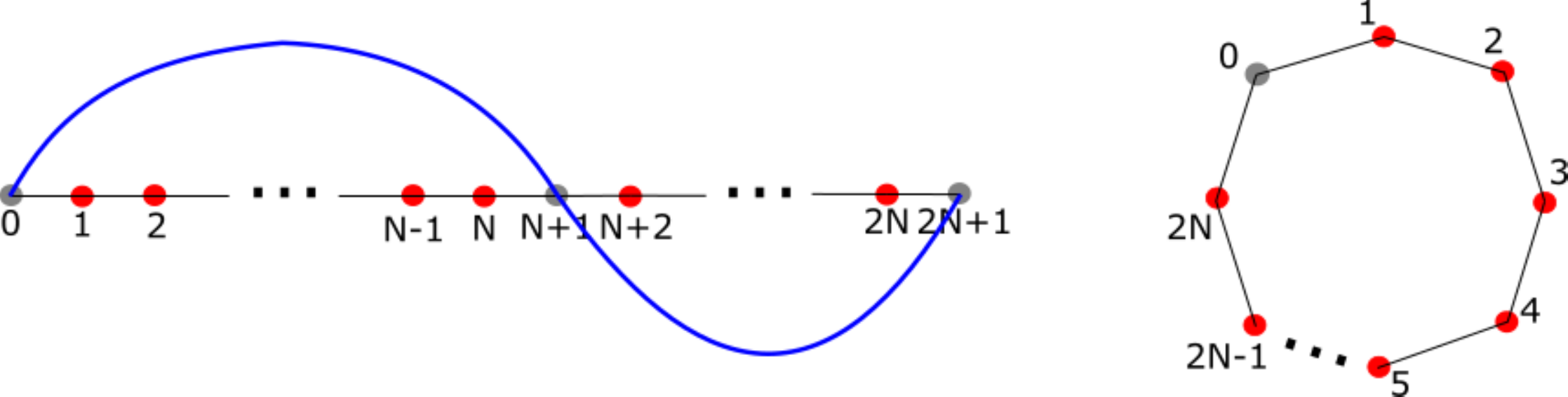}
	\caption{Pictorial representation of the lowest energy even parity standing wave and equivalent tight-binding $2N+1$ ring at the right.}
	\label{y}
\end{figure} 
that results in $k=\frac{2\pi}{N+1}m$. However, we should recall that every $k$ has a direct correspondence to a $k$ contained in the first Brillouin zone, meaning that they represent the same wavefunction. By taking a look at all $k$ contained in the first Brillouin zone it is clear that only half of the solutions are present, since every pair $k$ and $-k$ correspond to the same wavefunction solution, since $\ket{k,open}=-\ket{-k,open}$. Besides, $k=0$ is not a solution. So, until now, we have obtained the set of unique solutions $k=\frac{2\pi}{N+1}m^{'}$, where $m^{'}=1,..,\frac{N}{2}$ (we have excluded $k=0$ and assumed $N$ even) and the wave functions are

\begin{equation}
\ket{k,open}=\sqrt{\frac{2}{N+1}}\left(\sum_{j}\sin(kj)c^\dagger_j\right)\ket{\emptyset}
\end{equation}
Now, we take into account the even parity solutions (see Fig. \ref{y}). In this case, the standing wave will not repeat itself after it reaches the boundary, but would repeat itself were it to travel $2(N+1)$ sites. In this sense, the problem now corresponds to one of a tight-binding ring of $2(N+1)$ sites, with boundary conditions $\ket{0}=\ket{2(N+1)}$ and with the added condition that the amplitude of the wavefunction at site $N+1$ is zero, which will naturally be obeyed by any stationary wave of the form $\ket{k,open}$. We find the quantization by applying the boundary condition

\begin{equation}
\sin(k(2(N+1)))=2\pi m
,\qquad
n=0,...,2N+1
\end{equation}
that results in $k=\frac{2\pi}{2(N+1)}n$. However, same as above, half of the solutions are repeated, so we take those out, and $k$ becomes $k=\frac{2\pi}{2(N+1)}m^{'}$, where $m^{'}=1,...,N$.
Moreover, we have already took into consideration some of these $k$ values in the odd parity situation, so we take those out, which corresponds to taking out the even $m^{'}$, so we get $m^{'}=1,3,...,N-1$. The wavefunctions for the even parity standing waves will be

\begin{equation}
\ket{k,open}=\sqrt{\frac{2}{N+1}}\left(\sum_{j}\sin(kj)c^\dagger_j\right)\ket{\emptyset}.
\end{equation}
In Fig. \ref{energy_open} we plot the energy levels for a open chain with ten sites in units of $t$ and in Fig. \ref{ground_open} we plot the amplitude distribution of the ground state for the same chain.

\chapter{Kitaev Chain}
"Implementing a full-scale quantum computer is a major challenge to modern physics and engineering" \cite{kitaev2001unpaired}. One of the main problems is that it is very difficult to isolate quantum states from unwanted perturbations, and so the information being manipulated is subject to error. Although this problem is also recurrent in classical computation, error-correcting codes are enough to keep the information coherent. Quantum decoherence is the process of loosing information from a system to the environment and, to perform quantum computation, one must be able to preserve coherence of the states being manipulated. Several quantum error-correcting algorithms can be used to cover the effects of quantum decoherence. However, the problem might be solved at the physical level, by constructing decoherence protected degrees of freedom. Motivated by this, in this chapter we will study the Kitaev chain. The system is a simple toy model that exhibits unpaired Majorana fermions. It consists of a one-dimensional tight-binding chain, a quantum wire, that lies on the surface of a p-wave type superconductor, which induces a non standard superconducting term in the Hamiltonian, that couples only electrons with the same spin, so the problem effectively becomes one of spinless fermions, meaning that the site occupation can only be 0 or 1.
\section{Majorana Fermions}
\paragraph{Derivation}
We start by considering the following Hamiltonian that describes an $N$ site tight-binding chain of effectively spinless fermions with an induced superconducting pairing term,

\begin{equation}
H = -t\sum^{N-1}_{j=1}(c^\dagger_{j+1}c_{j}+c^\dagger_{j}c_{j+1})+\sum^{N-1}_{j=1}(\Delta c_{j}c_{j+1}+\Delta^* c^\dagger_{j+1}c^\dagger_{j})-\mu\sum^{N}_{j=1}c^\dagger_jc_j,
\end{equation}
where $t$ is the hopping parameter, $\mu$ is the chemical potential and $\Delta$ is the p-wave superconducting gap.
For the special case $t=\Delta$ and $\mu=0$, the Hamiltonian simplifies to

\begin{equation}
H=-t\sum^{N-1}_{j=1}\left( c^\dagger_jc_{j+1}+c^\dagger_{j+1}c_j-c^\dagger_{j+1}c^\dagger_{j}-c_jc_{j+1}\right).
\end{equation}
Kitaev proposed a method to diagonalize the Hamiltonian by defining the following Majorana operators \cite{kitaev2001unpaired}

\begin{equation}
	\gamma_{j,1}=c_j+c^\dagger_j
	,\qquad
	\gamma_{j,2}=i(c^\dagger_j-c_j).
\end{equation}
Each fermionic operator corresponds to a combination of Majorana operators,

\begin{figure}
	\centering
	\includegraphics[scale=0.7]{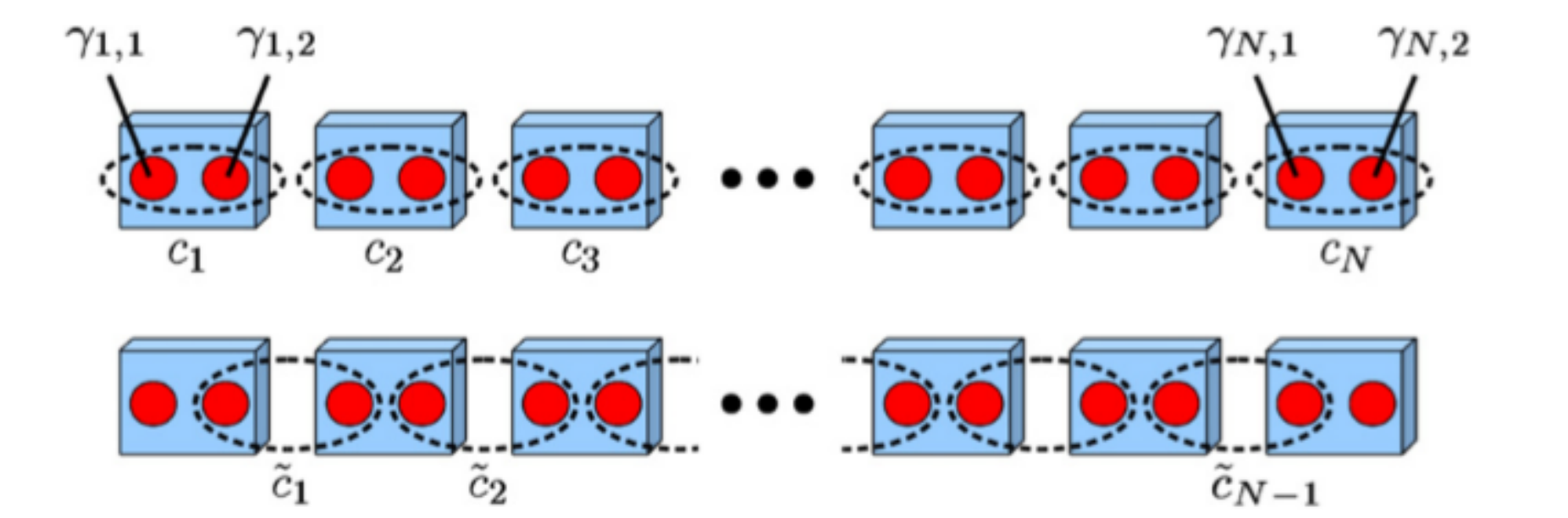}
	\caption{Pictorial representation of the Kitaev one-dimensional tight-binding chain with an induced p-wave superconducting term. In the limit $\mu=0$ and $t=\Delta$, the Hamiltonian is diagonalized in the basis of the fermionic operators obtained by combining Majorana operators from nearest-neighboring sites. This leaves to unpaired Majorana operators $\gamma_{N,2}$ and $\gamma_{1,1}$ at the ends. Extracted from \cite{leijnse2012introduction}.}
	\label{Majorana_figure}
\end{figure}

\begin{equation}
	c_j=\frac{1}{2}(\gamma_{j,1}+i\gamma_{j,2}),
	\qquad
	c_j^\dagger=\frac{1}{2}(\gamma_{j,1}-i\gamma_{j,2})
	\label{fermionic_majorana_relation}.
\end{equation}
To an unpaired Majorana operator, i.e., one decoupled from the system, one calls a Majorana fermion. Majorana operators correspond to particles which are their own anti-particle ($\gamma^\dagger\equiv\gamma$) and obey the following anti-commutation relation,

\begin{equation}
	\{\gamma_{i,\alpha},\gamma_{j,\beta}\}=2\delta_{ij}\delta_{\alpha\beta}.
	\label{anticommutation_majorana}
\end{equation}
Using \eqref{fermionic_majorana_relation} and the usual fermionic anti-commutation relations

\begin{equation} 
\{c_i,c^\dagger_j\}=\delta_{ij},
\qquad
\{c_i,c_j\}=\{c^\dagger_i,c^\dagger_j\}=0,
\end{equation}
one arrives at the following Hamiltonian, written in the basis of Majorana operators,

\begin{equation}
H=it\sum^{N-1}_{i=1}\gamma_{i,2}\gamma_{i+1,1}.
\label{H_majorana}
\end{equation}
Now we perform another change in basis by defining new fermionic operators $\tilde{c}_j$ that are a linear combinations of Majorana operators from nearest-neighboring sites, as shown at the bottom of Fig. \ref{Majorana_figure} ,

\begin{equation}
\tilde{c}_j=\frac{1}{2}(\gamma_{j+1,1}+i\gamma_{j,2})
,\qquad
\tilde{c}_j^\dagger=\frac{1}{2}(\gamma_{j+1,1}-i\gamma_{j,2}).
\end{equation}
The following inverse relations hold

\begin{equation}
\gamma_{j+1,1}=\tilde{c}^\dagger_j+\tilde{c}_j
,\qquad \gamma_{j,2}=i(\tilde{c}^\dagger_j-\tilde{c}_j).
\label{majorana_newfermionic_relation}
\end{equation}
In the basis of these new fermionic operators, the Hamiltonian becomes diagonalized,

\begin{equation}
H=it\sum^{N-1}_{j=1}\gamma_{j,2}\gamma_{j+1,1}=-2t\sum^{N-1}_{j=1}(\tilde{c}^\dagger_j\tilde{c}_j-\frac{1}{2}).
\label{eq_5}
\end{equation}
These new fermionic operators represent quasiparticles of the system with energy $-2t$. We notice that we only need $(N-1)$ quasiparticle fermionic operators to write the Hamiltonian in this new basis, so one more state must exist. This state can be constructed from the unpaired Majorana operators $\gamma_{N,2}$ and $\gamma_{1,1}$, which are missing from the Hamiltonian in \eqref{H_majorana} and, since they are decoupled from the system, they combine to create a Majorana state of zero energy. In the quasiparticle fermionic basis, these Majorana operators can be written as

\begin{equation}
\tilde{c}_M=\frac{1}{2}(\gamma_{N,2}+i\gamma_{1,1}),
\qquad
\tilde{c}^{\dagger}_M=\frac{1}{2}(\gamma_{N,2}-i\gamma_{1,1}).
\label{majorana_operators}
\end{equation}
The Majorana state given  by $c_M^\dagger$ is a highly non-local state with finite amplitude only at the edges of the chain. Due to its non-local nature, this state is immune to perturbations.

We may now look at the problem in a different way, where we have $N-1$ sites, which can be occupied by quasiparticles, each adding $-2t$ energy when occupied and another extra site, which can be occupied by the fermion defined in equation \eqref{majorana_operators}, with no energetic cost. To fully characterize the state of the system, we must specify the action of every quasiparticle creation operator, including $\tilde{c}^{\dagger}_M$, on the vacuum of the system. Since the operators in equation \eqref{majorana_operators} do not show up in the Hamiltonian, hence do not affect the energy of the state on which they act, every energy should always be, at least, doubly degenerate.

The above argument has only been made for a very special case. However, one can show that Majorana end states still exist as long as the condition $|\mu|<2t$ is verified \cite{leijnse2012introduction}. In the general case, Majorana end states are not completely localized at the edges of the chain and exhibit a decaying tail to the bulk.
\paragraph{Computational Results}
We will now, very briefly, show some numerical results obtained using Matlab for the Kitaev chain. We will show a plot of the energy levels and energy dispersion relation with $t$ for the $N=3$, $N=4$ and $N=5$ Kitaev chain and draw simple conclusions from them. The simulations are made for $\Delta=1$ and $\mu=0$.

\begin{figure}
	\centering
	\begin{minipage}{0.4\textwidth}
	\centering
\includegraphics[scale=0.4]{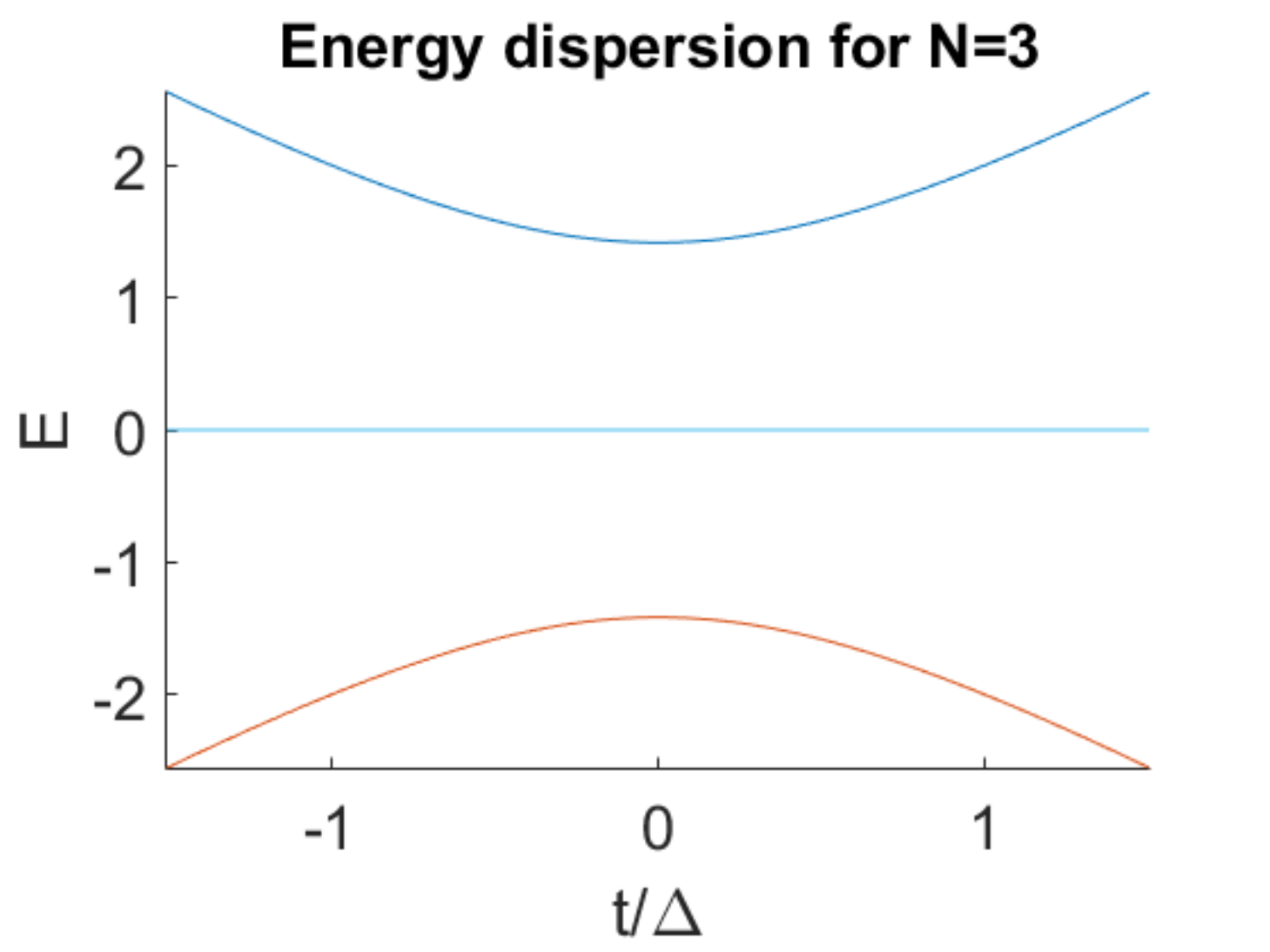}
\caption{Energy levels of the Kitaev Hamiltonian in \eqref{eq_5} as a function of $t$ for $N=3$. Energy is in units of $\Delta$.}
\label{figura2}
	\end{minipage}%
~ ~
	\begin{minipage}{0.4\textwidth}
	\centering
\includegraphics[scale=0.4]{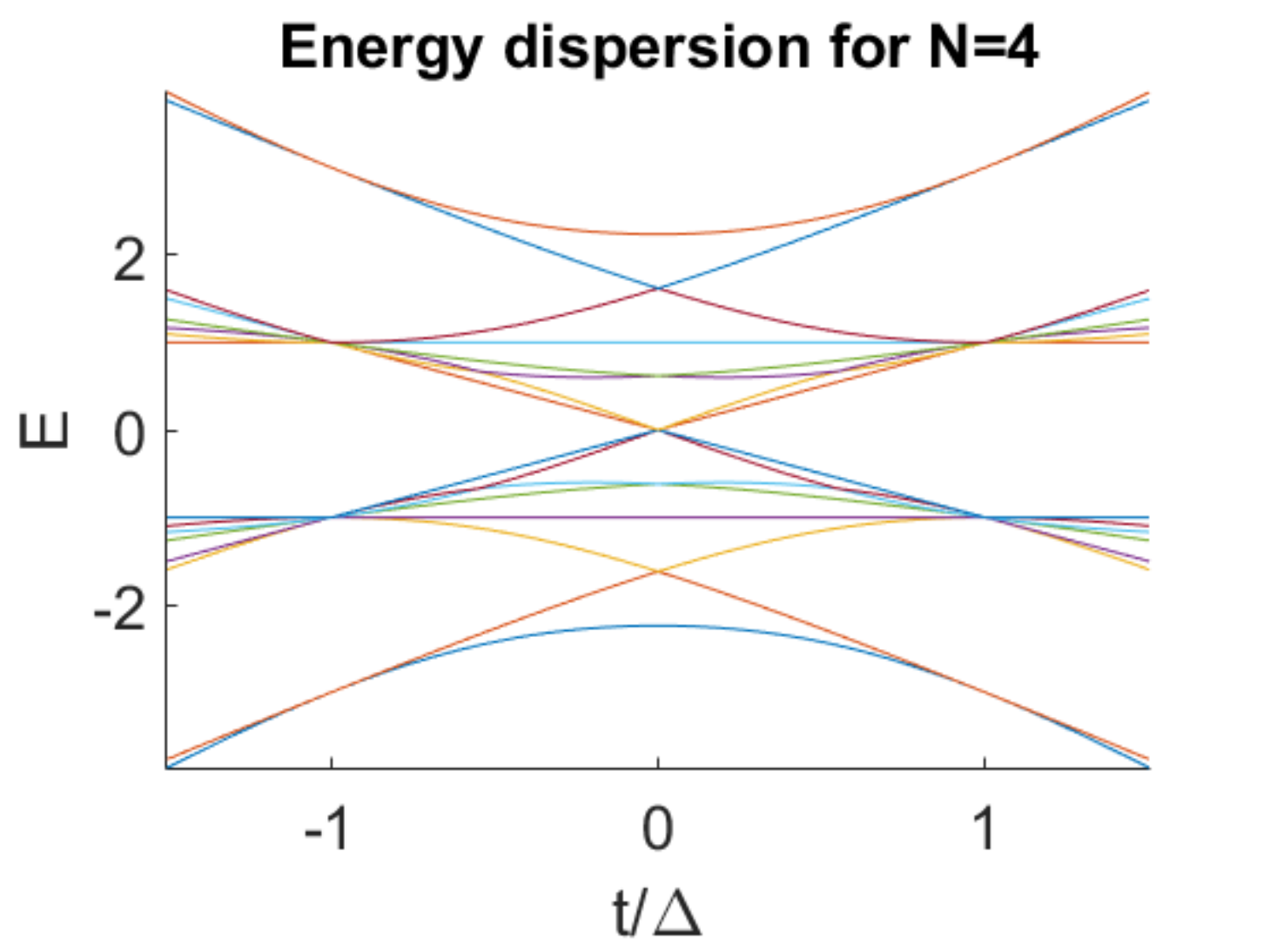}
\caption{Energy levels of the Kitaev Hamiltonian in \eqref{eq_5} as a function of $t$ for $N=4$. Energy is in units of $\Delta$.}
\label{figura11}
	\end{minipage}
\end{figure}

\begin{figure}
	\centering
	\includegraphics[scale=0.5]{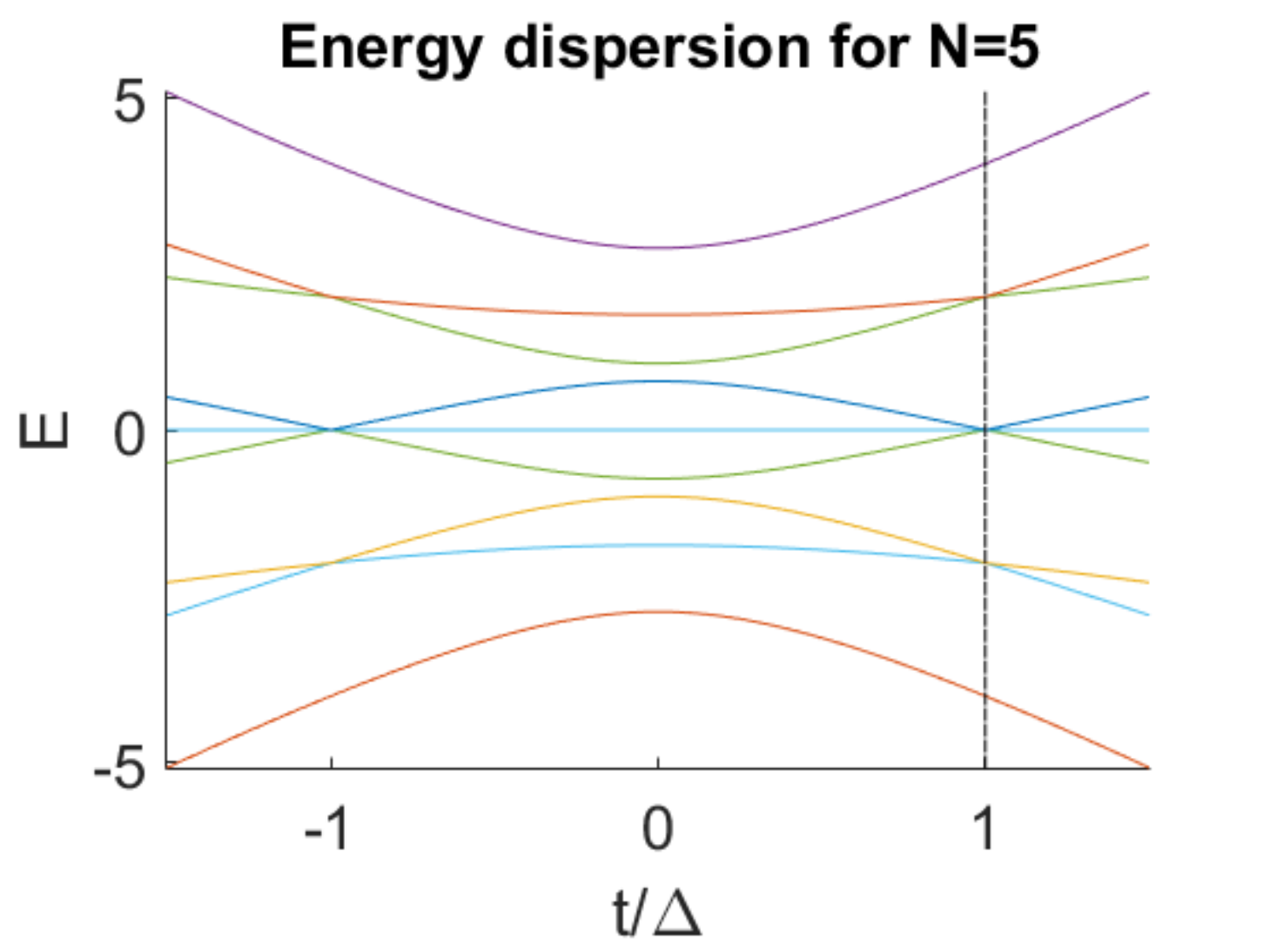}
	\caption{Energy levels of the Kitaev Hamiltonian in \eqref{eq_5} as a function of $t$ for $N=5$. Energy is in units of $\Delta$. The dashed line marks the regime $t=\Delta$.}
	\label{figura3}
\end{figure}

\begin{figure}[!tbp]
	\centering
	\subfloat[ ]{\includegraphics[width=0.33\textwidth]{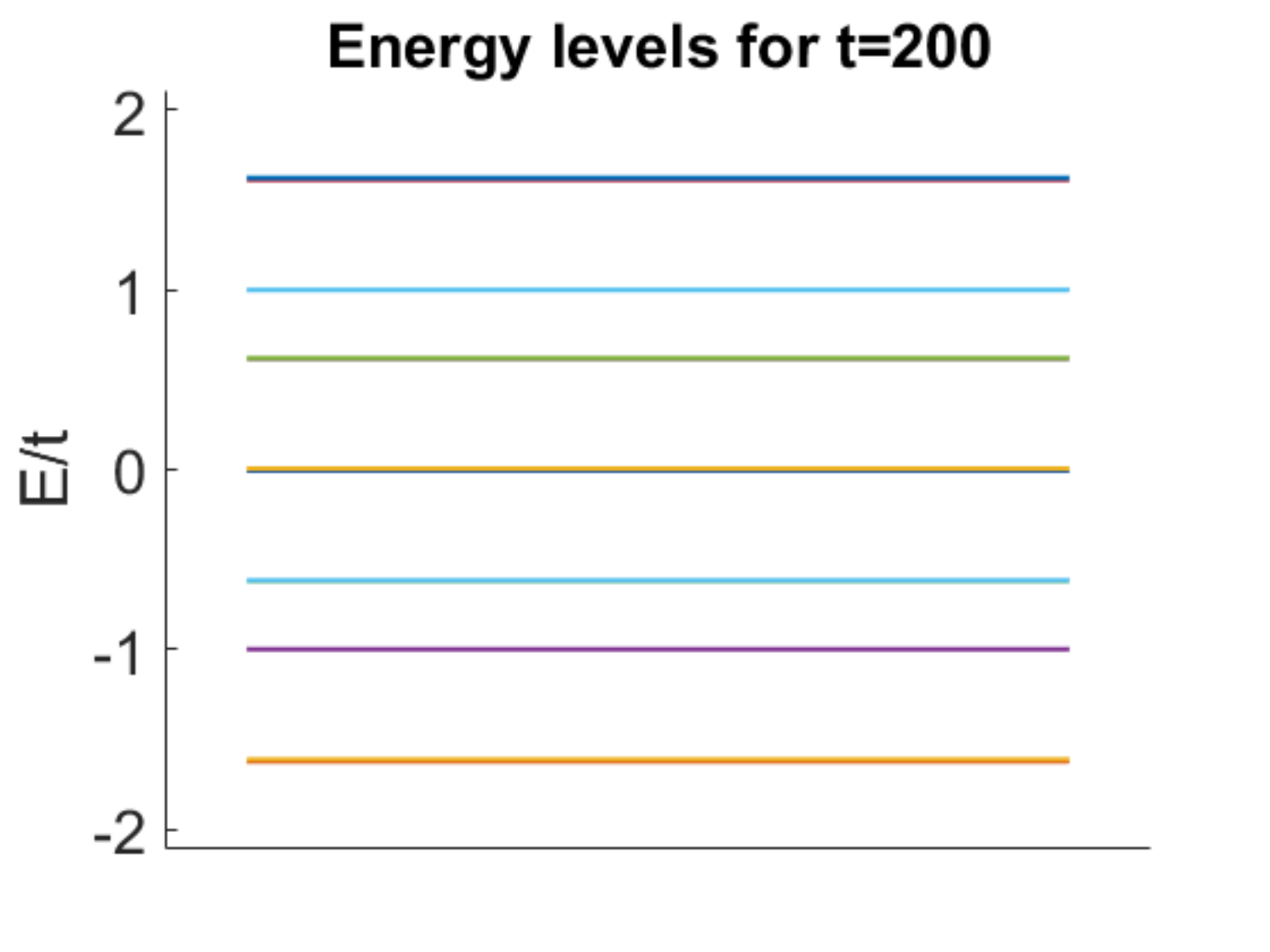}\label{Energy_Kitaev_200}}
	\subfloat[ ]{\includegraphics[width=0.33\textwidth]{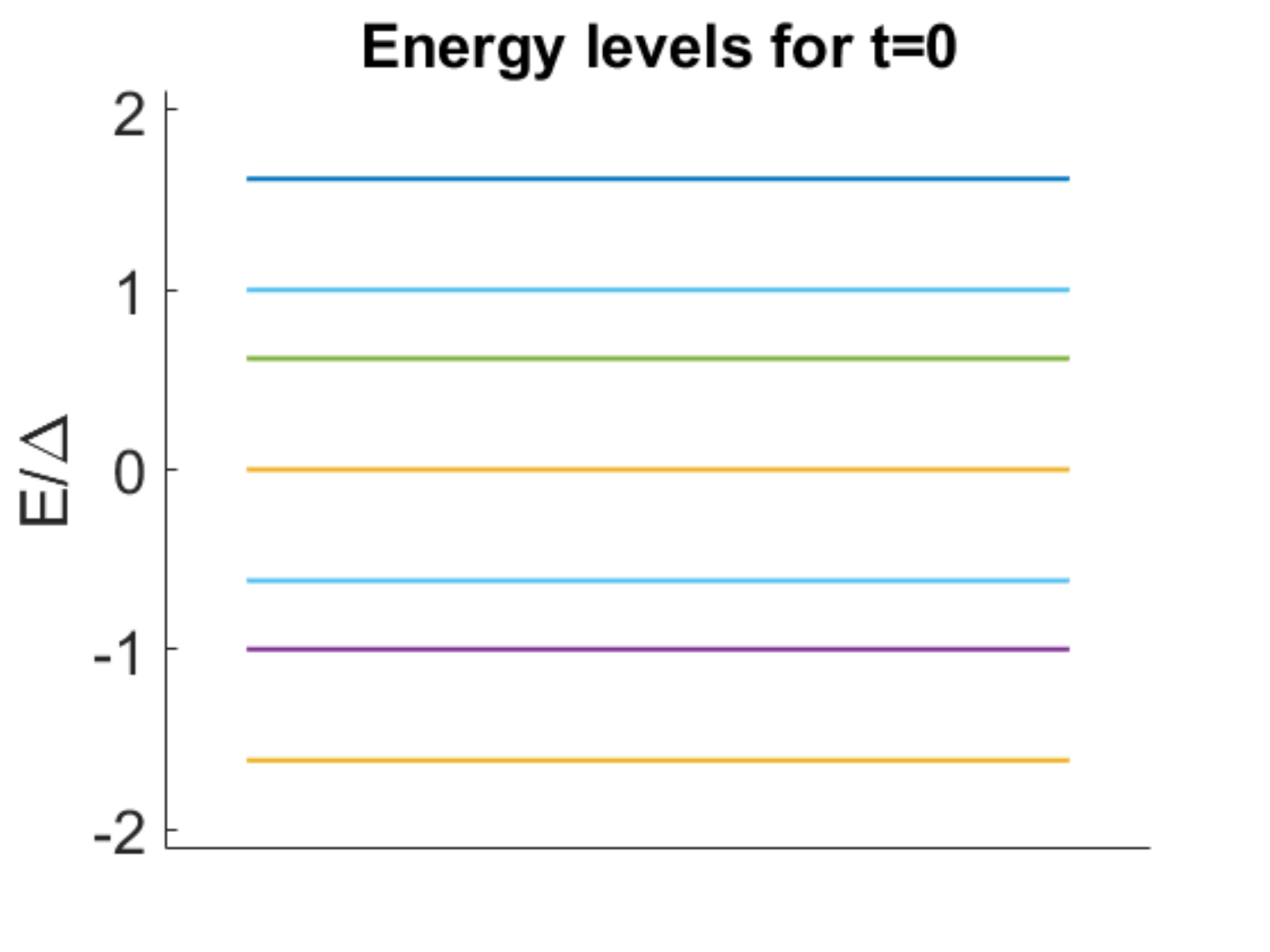}\label{Energy_Kitaev_0}}
	\caption{Energy levels in units of $t$ for (a) $t=200\Delta$ ($t\rightarrow\infty$) and in units of $\Delta$ for (b) $t=0$. We cn see that both these regimes yield almost the same energy spectrum.}
	\end{figure}

\begin{figure}[!tbp]
	\centering
	\subfloat[ ]{\includegraphics[width=1\textwidth]{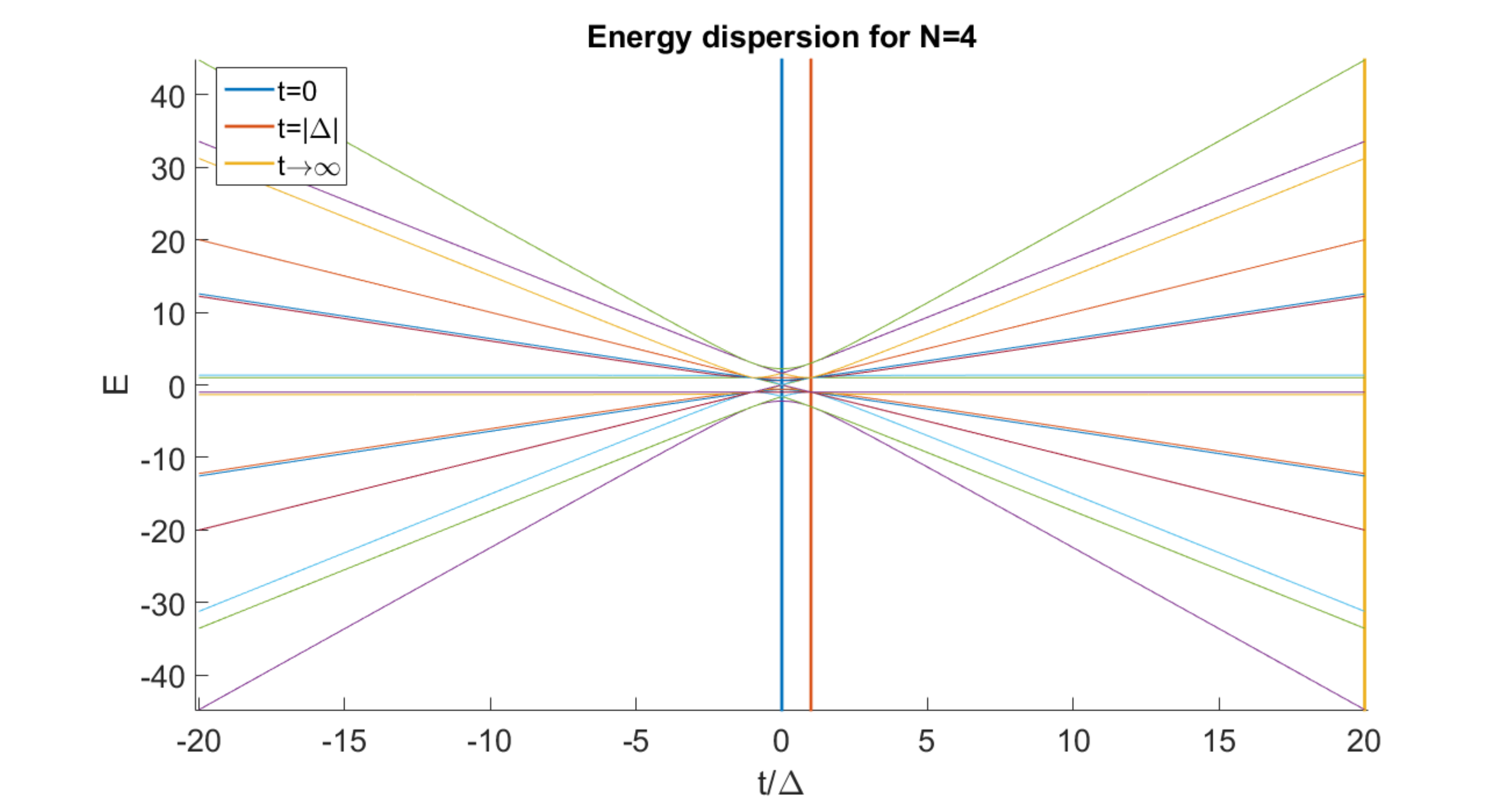}\label{tretas}}

	\subfloat[ ]{\includegraphics[width=1\textwidth]{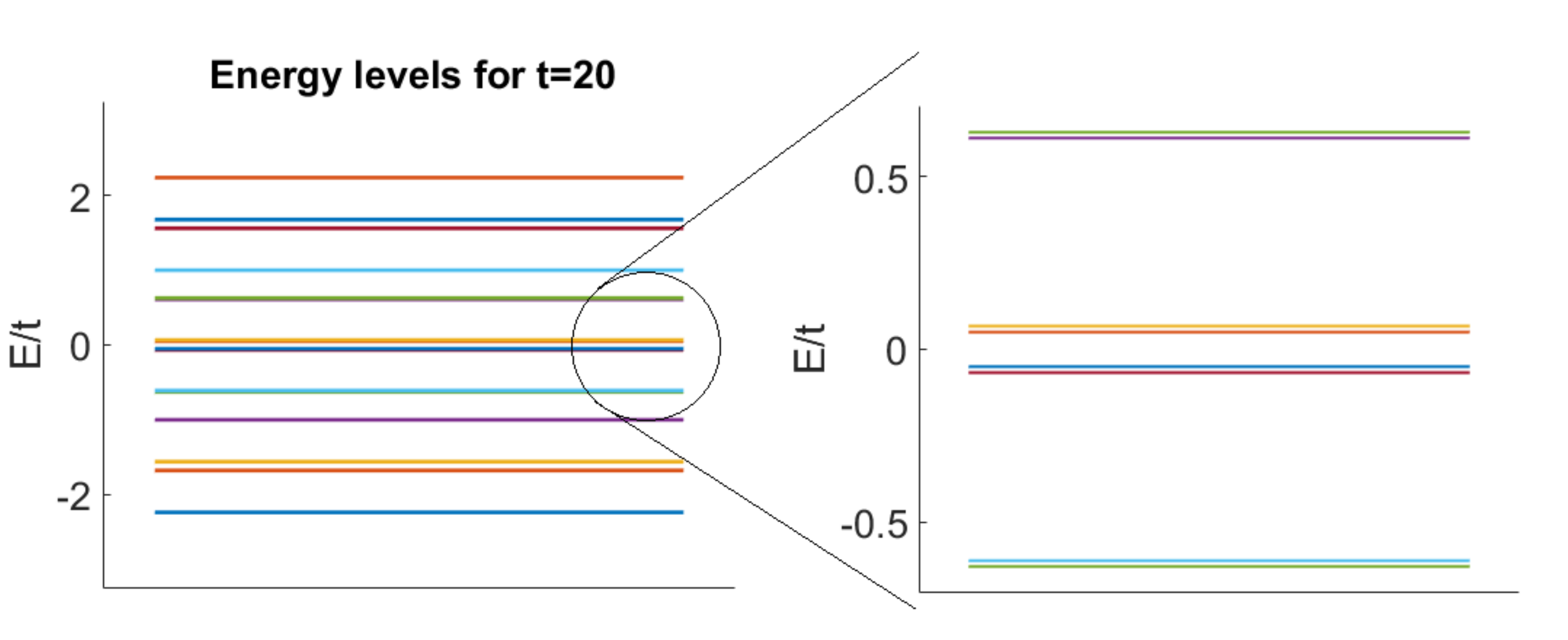}\label{tretas2}}
	\caption{(a) Energy levels as a function of $\frac{t}{\Delta}$ (in units of $\Delta$) for $N=4$. The vertical lines mark the three regimes $t=0$, $t=|\Delta|$ and $t=20\Delta$ (approaching the limit $t\rightarrow\infty$). (b) Energy levels in units of $t$ when $t=20\Delta$.}
\end{figure}

\begin{figure}[!tbp]
	\centering
	\subfloat[ ]{\includegraphics[width=0.47\textwidth]{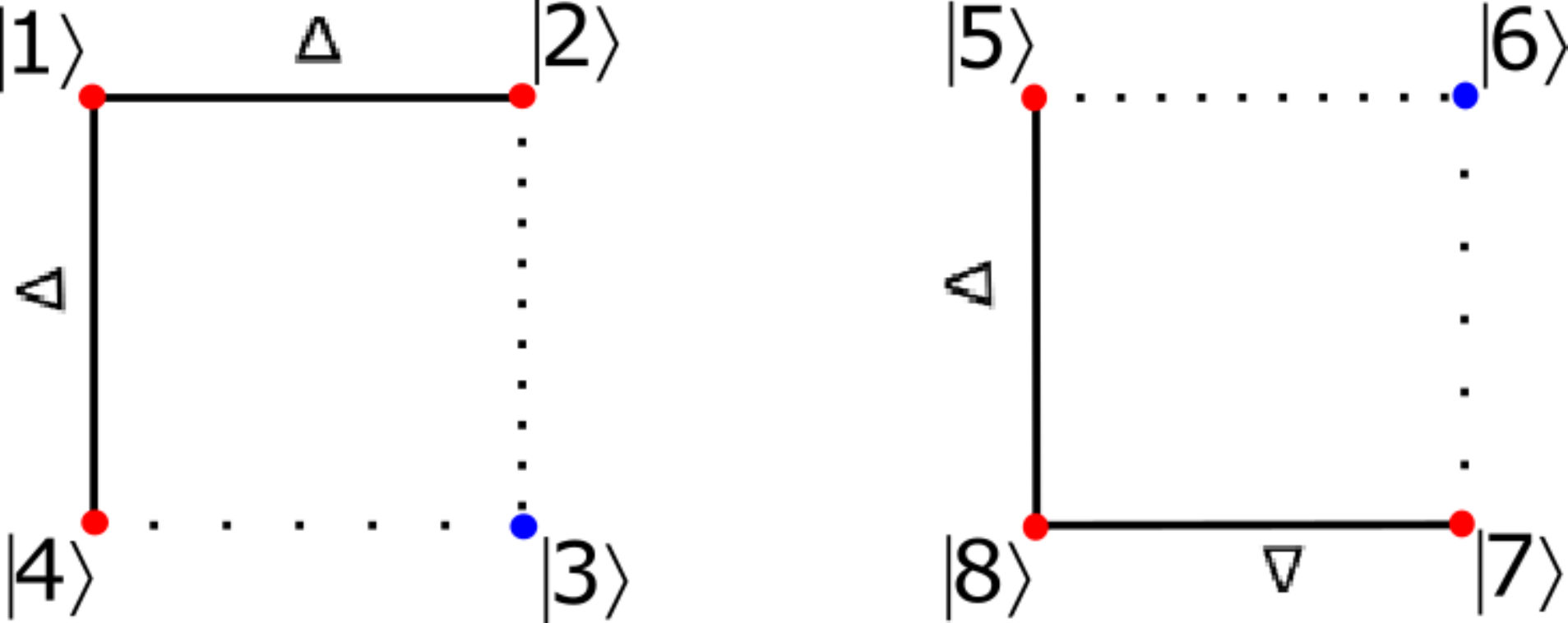}\label{Q_1}}
	\hfill
	\subfloat[ ]{\includegraphics[width=0.47\textwidth]{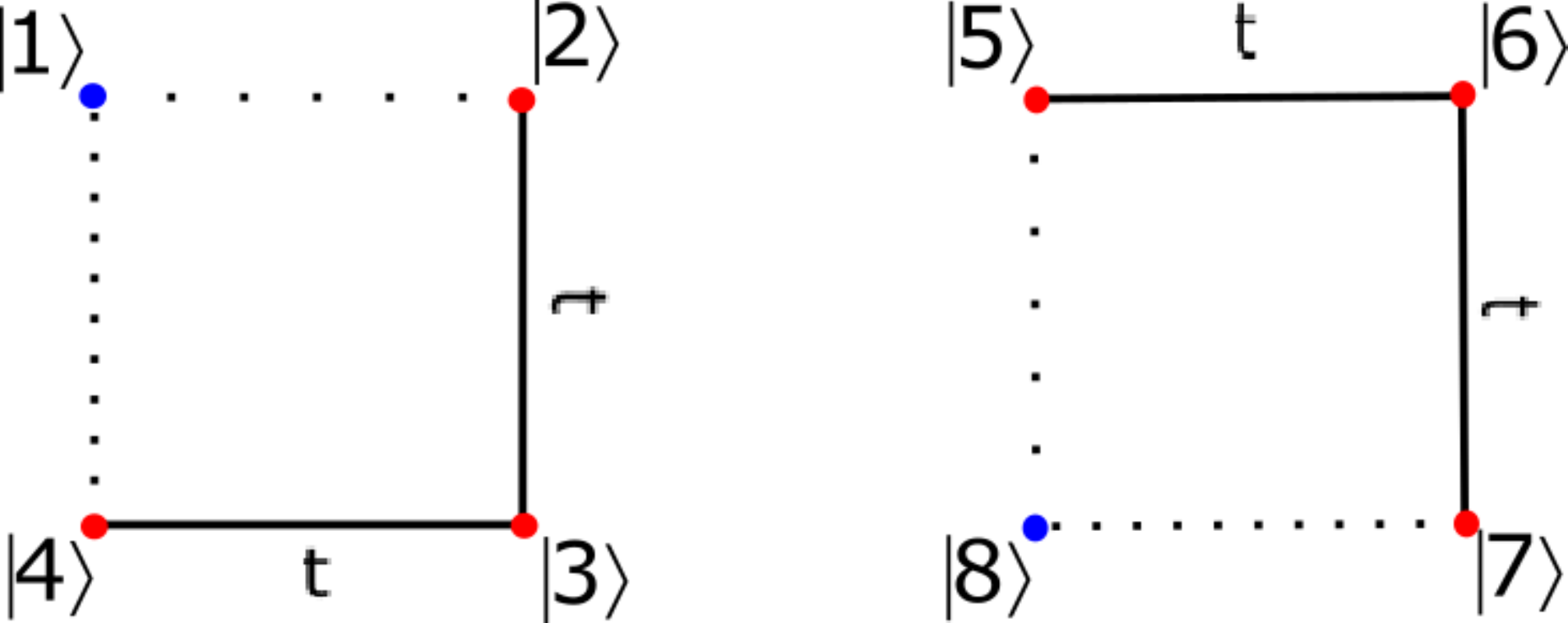}\label{Q_2}}
	\caption{Tight-binding lattices correspondent to the three site Kitaev chain for the different regimes (a) $t=0$ and (b) $t\rightarrow\infty$. The sites in blue are isolated.}
\end{figure}

By looking at the energy levels for the three chains of Fig. \ref{figura2}, Fig. \ref{figura11}, and Fig. \ref{figura3}, we can distinguish between three regimes where we know the exact solution of the problem. One regime, already analyzed in this section, is where $t=\Delta$ and $\mu=0$. We notice that, in this regime, for any $N$, the energy degeneracy reaches its maximum, being equal to the number of sites of the chain. This is to be expected if we look at the Hamiltonian in equation \eqref{eq_5} since by adding one $\tilde{c}_j^\dagger$ quasiparticle the energy of the system decreases by $2t$ and there only exists $N-1$ quasiparticle operators for a chain with $N$ sites, so one can only decrease the energy $N-1$ times. Besides, in this regime, as will be studied from section \ref{threesite} on, the system as a correspondence between two independent and identical tight-binding lattices with $2^{N-1}$ sites. 

The other regime occurs for $t=0$. Here we can conclude that the system will correspond to a tight-binding problem constituted of two independent lattices, by the same reasoning as in section \ref{threesite}. The lattices will have the shape of Fig. \ref{Q_1} for $N=3$. The correspondent tight-binding lattice would be the same were we to analyze the regime $\Delta\rightarrow\infty$. 

Finally, the last regime occurs when $t\rightarrow\infty$ and in this case the correspondent tight-binding lattice will have the form of Fig. \ref{Q_2} for $N=3$. The correspondent tight-binding lattice would be the same were we looking at the regime $\Delta=0$. In Fig. \ref{tretas}, we plot the energy spectrum for $N=4$ as a function of $\frac{t}{\Delta}$, where the different regimes are indicated by the vertical lines, and in Fig. \ref{tretas2}, we plot the energy levels in units of $t$ for the case $t=20\Delta$. We verify that the energy levels in units of $t$ when $t\rightarrow\infty$ tend to the energy levels in units of $\Delta$ when $t=0$ (compare Fig. \ref{Energy_Kitaev_200} with Fig. \ref{Energy_Kitaev_0}), which is expected because the lattices in Fig. \ref{Q_1} and Fig. \ref{Q_2} should have the same energy levels.

For every regime, it is straightforward to find the eigenfunctions and eigenvalues of the Kitaev chain because of the correspondence with simple tight-binding systems. In the following sections, we analyze the Kitaev chain and study the first regime ($t=|\Delta|$), providing the rules of construction for the correspondent tight-binding lattices for any $N$.

\newpage

\section{Band Structure of the Kitaev chain}
In this section, we will calculate the band structure of the bulk of the Kitaev chain. We start by considering the full Kitaev Hamiltonian with periodic boundary conditions,

\begin{figure}[!tbp]
	\centering
	\subfloat[ ]{\includegraphics[width=0.33\textwidth]{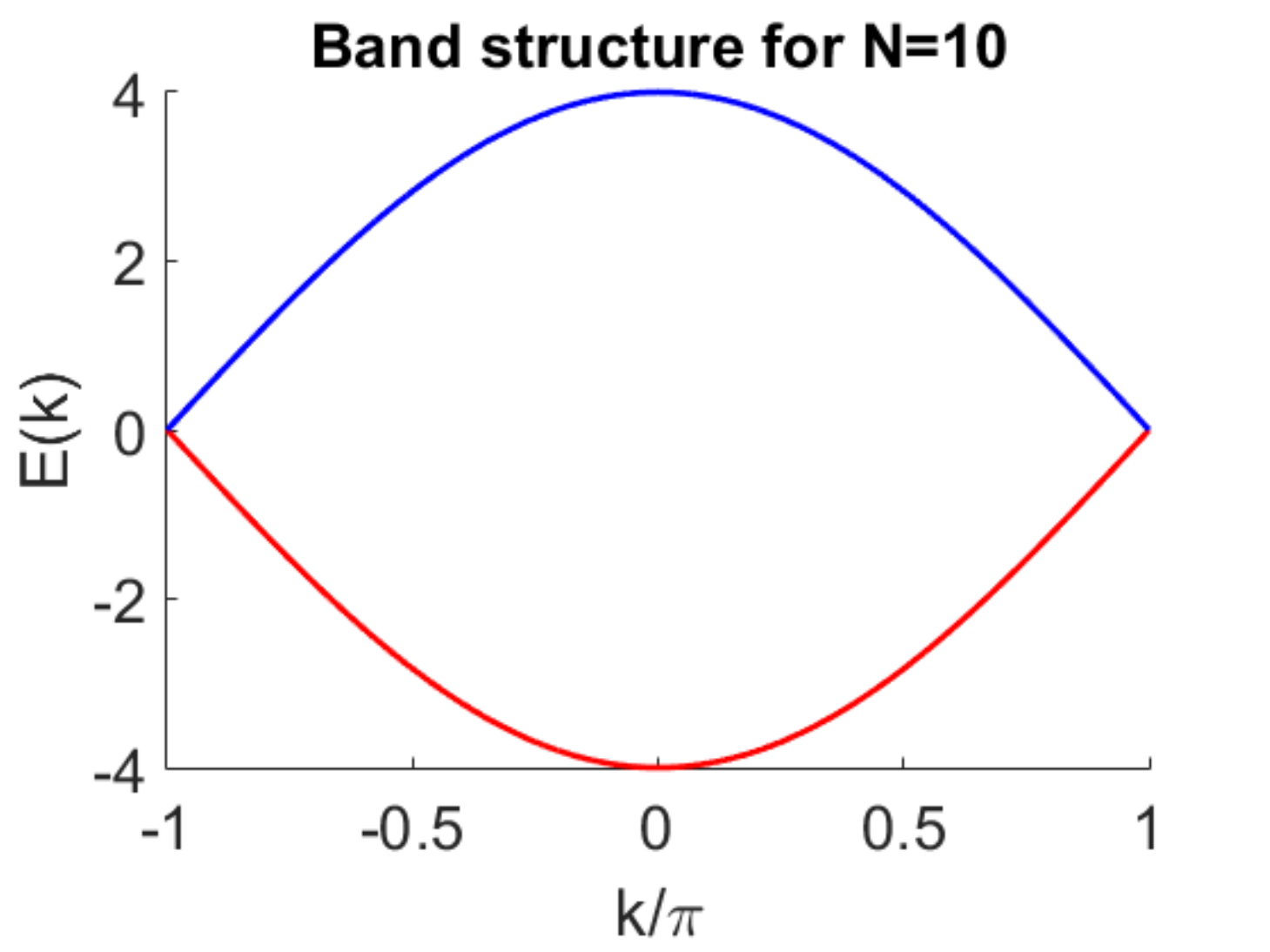}\label{Kitaev_Band_1}}
	\hfill
	\subfloat[ ]{\includegraphics[width=0.33\textwidth]{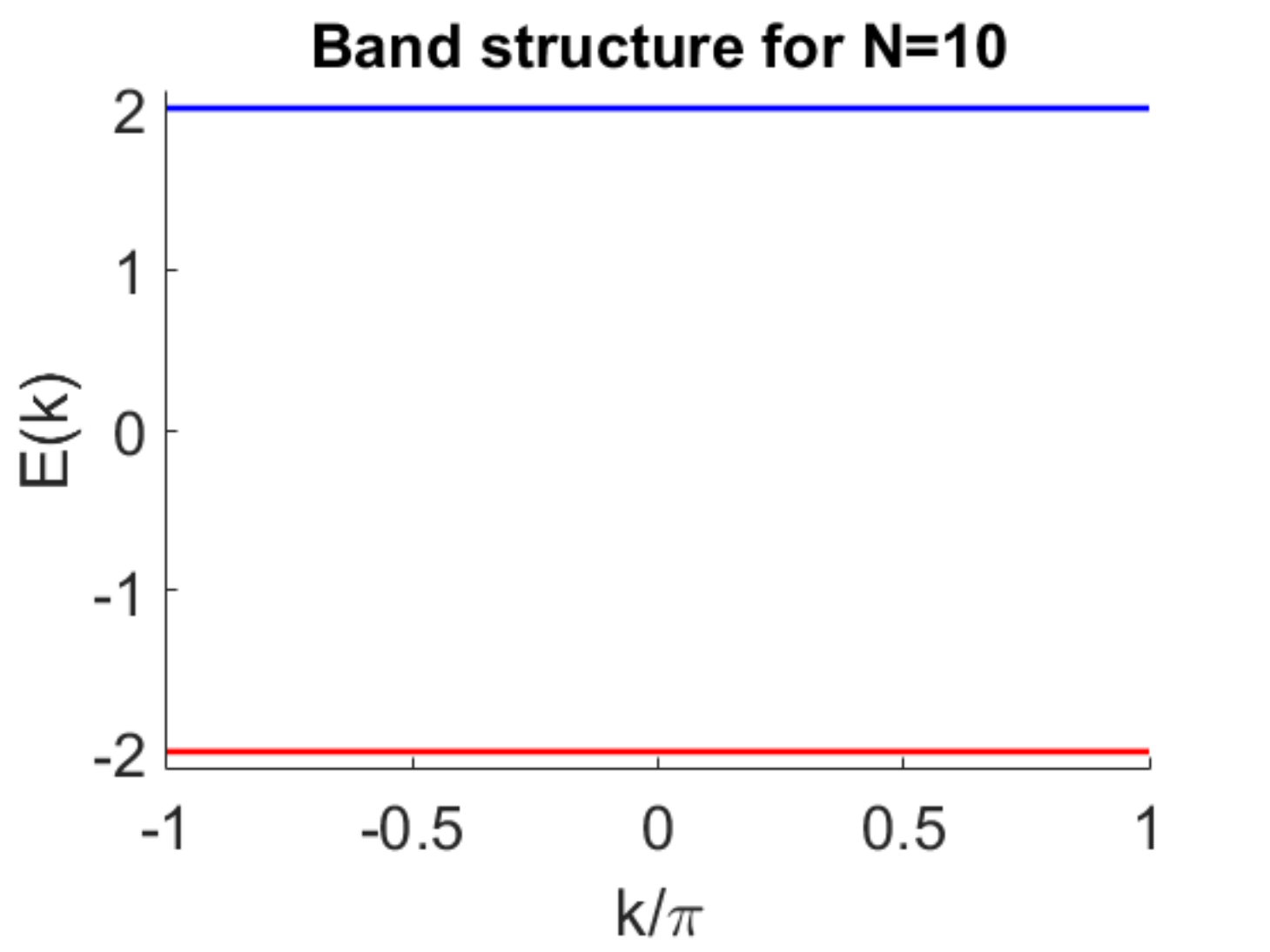}\label{Kitaev_Band_3}}
	\hfill
	\subfloat[ ]{\includegraphics[width=0.33\textwidth]{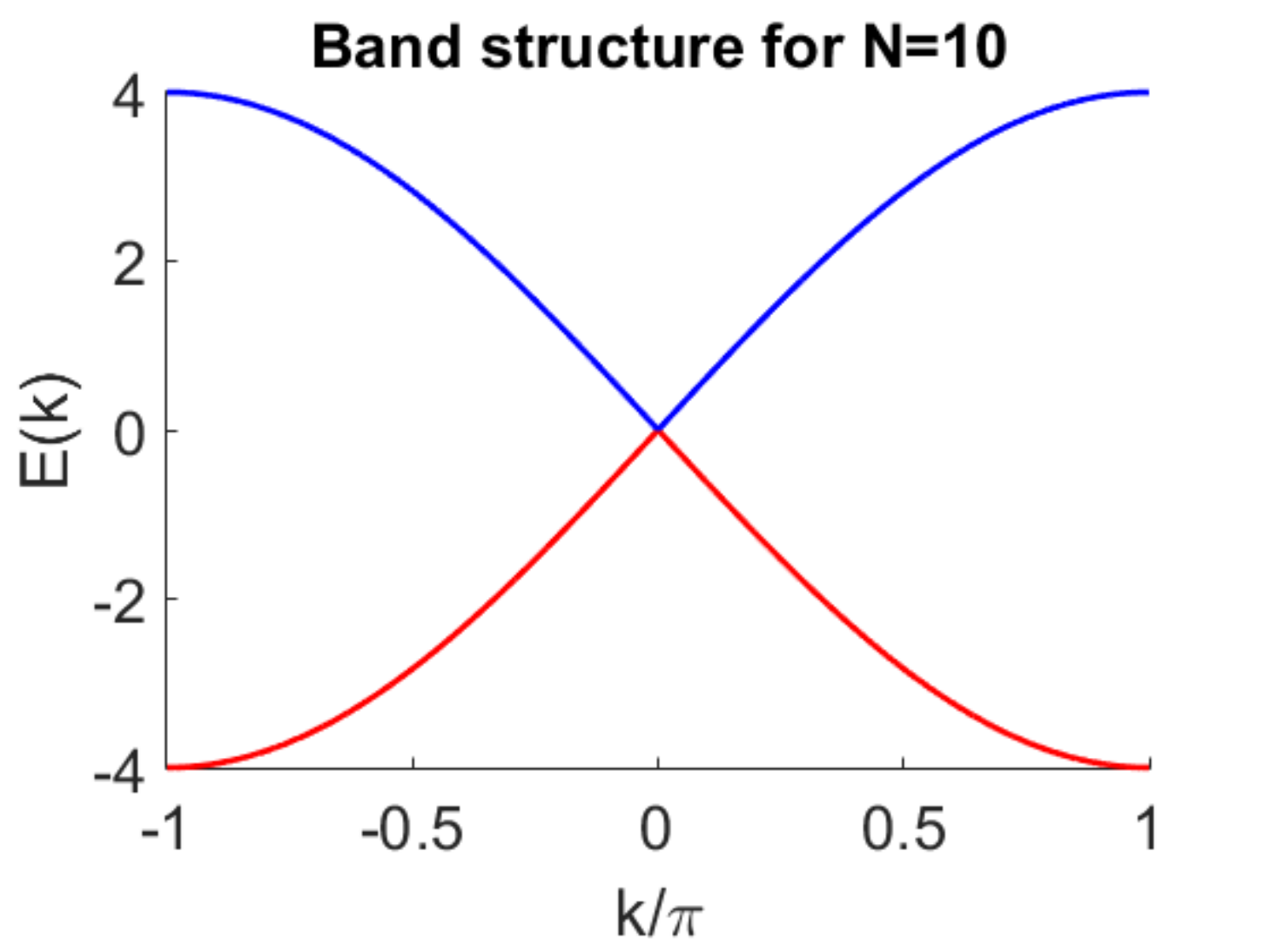}\label{Kitaev_Band_2}}
	\caption{Band structure of a ten site Kitaev chain with $t=\Delta$ for (a) $\mu=2t$, (b) $\mu=0$, (c) $\mu=-2t$. Energy is in units of $t$. The gap closes when the system reaches the limits of the topological phases $|\mu|=2t$. For $|\mu|>2t$ the system is gapped again but becomes topologically trivial, that is, Majorana states are no longer present in the Kitaev chain.}
\end{figure}

\begin{equation}
H =\sum^{N-1}_{j=1}\left( -tc_j^\dagger c_{j+1}+\Delta c_{j+1}^{\dagger}c_j^\dagger +H.c.\right)-\mu\sum_{j=1}^{N}c_j^\dagger c_j.
\end{equation}
Using equations \eqref{Bloch_j} and the orthogonality relation \eqref{eq_orto}, we can write the Hamiltonian in $k$-space,

\begin{equation}
H=\sum_k \left(-te^{ik}c^\dagger_k c_{k}+\Delta e^{-ik}c^\dagger_k c^\dagger_{-k}+H.c\right)-\mu\sum_k c^\dagger_k c_{k}.
\end{equation}
We can avoid double counting and further simplify the result by restricting the sum to $k>0$,

\begin{eqnarray}
&\sum_k& -te^{ik} c^\dagger_k c_{k}=\sum_{k>0} -te^{ik}c^\dagger_k c_{k}-te^{-ik}c^\dagger_{-k} c_{-k}-tc^\dagger_{k=0} c_{k=0}= \nonumber\\
&\sum_{k>0}& -te^{ik}c^\dagger_k c_{k}-te^{-ik}(1-c_{-k}c^\dagger_{-k})=		\sum_{k>0} -te^{ik}c^\dagger_k c_{k}+te^{-ik}c_{-k}c^\dagger_{-k},
\\\nonumber
\\
&\sum_k& \Delta e^{-ik} c^\dagger_k c_{-k}^\dagger=\sum_{k>0} \Delta e^{-ik}c^\dagger_k c_{-k}^\dagger+\Delta e^{ik}c^\dagger_{-k} c_{k}^\dagger+\Delta c^\dagger_{k=0} c^\dagger_{k=0}= \nonumber\\
&\sum_{k>0}& \Delta e^{-ik}c^\dagger_k c_{-k}^\dagger+\Delta e^{ik}(1-c^\dagger_k c_{-k}^\dagger)=\sum_{k>0} -2i\Delta \sin(k)c^\dagger_k c_{-k}^\dagger,
\end{eqnarray}
where we dropped the surface term at $k=0$ and any other constant terms. Analogously, we restrict the sum in the other terms of the Hamiltonian and we get

\begin{equation}
H=\sum_{k>0}\begin{pmatrix}c_k^\dagger&c_{-k}\end{pmatrix}\begin{pmatrix}-2t\cos(k)-\mu&-2i\Delta \sin(k)\\
	2i\Delta \sin(k) & 2t\cos(k)+\mu\end{pmatrix}\begin{pmatrix}c_k\\
	c_{-k}^\dagger\end{pmatrix}.
\end{equation}
The only thing left to find the band structure is to diagonalize the Hamiltonian above. In the diagonal basis we get

\small
\begin{equation}
H=\sum_{k>0}\begin{pmatrix}a_k^\dagger&b_{k}^\dagger\end{pmatrix}\begin{pmatrix}-\sqrt{(2tcos(k)+\mu)^2+4\Delta^2sin^2(k)}&0\\
0 & \sqrt{(2tcos(k)+\mu)^2+4\Delta^2sin^2(k)}\end{pmatrix}\begin{pmatrix}a_k\\
b_{k}\end{pmatrix},
\end{equation}
\normalsize
where the operators $a_k^\dagger$ and $b_k^\dagger$ are linear combinations of $c_k^\dagger$ and $c_{-k}$ given by

\begin{equation}
	a_k^\dagger=\frac{1}{\sqrt{2}}(c_k^\dagger+e^{i\phi_1(k)}c_{-k})
	,\qquad
	b_k^\dagger=\frac{1}{\sqrt{2}}(c_k^\dagger+e^{i\phi_2(k)}c_{-k}),
\end{equation}
where $\phi_1(k)$ and $\phi_2(k)$ are phases with an elaborate dependence on $k$. In Fig. \ref{Kitaev_Band_1} and Fig. \ref{Kitaev_Band_2} we plot the band structure in units of $t$ for the case $\mu=2t$ and $\mu=-2t$ respectively.

In the regime where $t=|\Delta|$ we can calculate Zak's phase, which is a topological invariant, that will allow us to characterize if the system is in a topological trivial or non-trivial phase. Zak's phase is defined by 

\begin{equation}
	\Gamma_\beta=i\int_{-\pi}^\pi dk\braket{\beta(k)|\frac{d}{dk}\beta(k)}
\end{equation}
where $\ket{\beta(k)}=\beta_k^\dagger\ket{0}$, with $\beta=a,b$. We get the following values

\begin{equation}
	\left\{\begin{matrix}
	\Gamma_{a}=\Gamma_{b}=0,\qquad\text{for $|\mu|>2t$},\\
	\Gamma_{a}=\Gamma_{b}=\pi,\qquad\text{for $|\mu|<2t$}
	\end{matrix}\right..
\end{equation}
The $\pi$ shift between these two $\mu$ regimes is indicative of a transition from a topological trivial to a non-trivial phase.

\section{Kitaev $\Leftrightarrow$ Tight-binding correspondence for a three-site chain \label{threesite}}
A property of the system we have been studying is that it has a correspondence to a periodic tight-binding one-dimensional lattice. In order to establish this correspondence, we will consider a Kitaev chain of only three sites. We will then generalize for any number of sites $N$ of the Kitaev chain.

\begin{figure}
	\centering
	\includegraphics[scale=0.5]{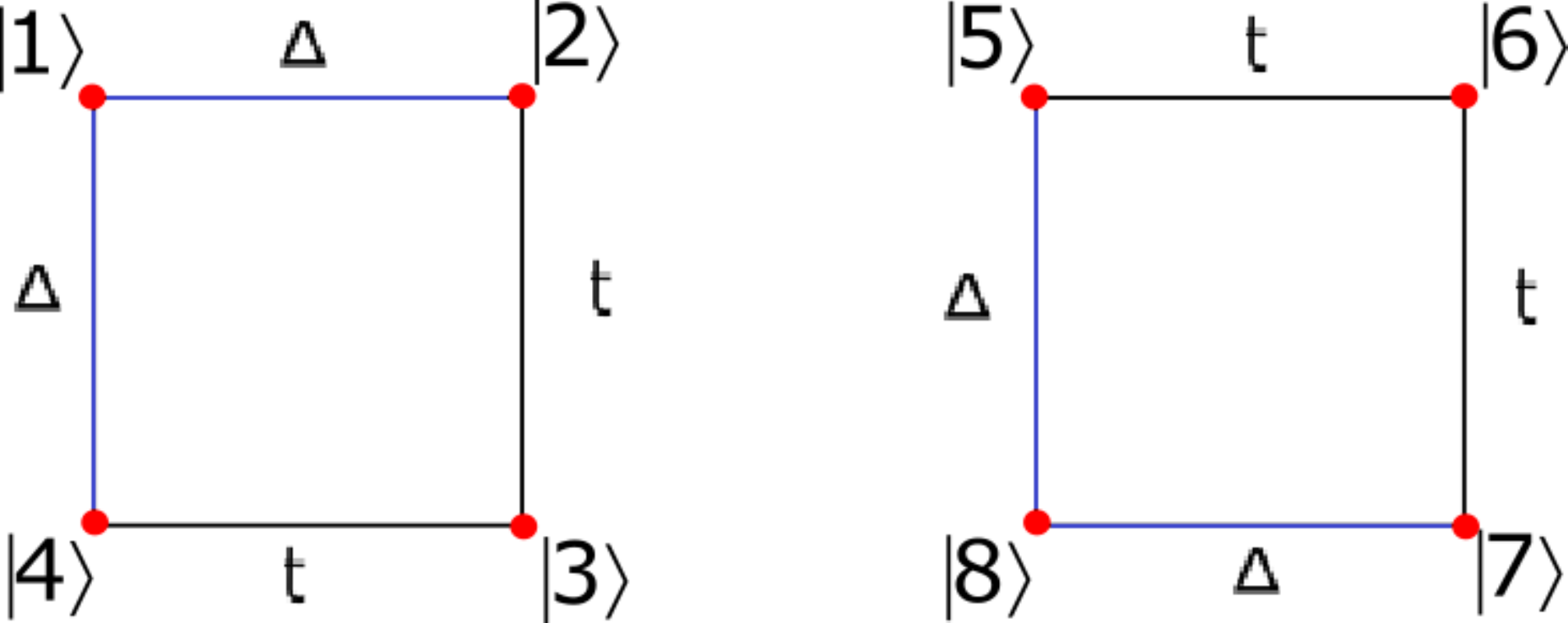}
	\caption{Tight-binding lattices correspondent to the three-sites Kitaev chain}
	\label{TB_4}
\end{figure}

We consider the Hamiltonian of the Kitaev chain for $N=3$ and $\mu=0$,

\begin{equation}
	H = -t\sum^{2}_{j=1}c^\dagger_{j+1}c_{j}+H.c.+\Delta\sum^{2}_{j=1}c_jc_{j+1}+H.c.
\end{equation}
The set of states $B=\{\ket{000},\ket{110},\ket{101},\ket{011},\ket{100},\ket{010},\ket{001},\ket{111}\}$ constitutes an orthogonal basis in which we can express the Hamiltonian\footnote{We used the notation where, for example, the state $\ket{111}$ is $c_1^\dagger c_2^\dagger c_3^\dagger\ket{000}$, where $\ket{000}$ is the vacuum.}. In matrix form the Hamiltonian is given by

\begin{equation}
H=\begin{pmatrix}
0 & -\Delta & 0 & -\Delta & 0 & 0 & 0 & 0\\
-\Delta^* & 0 & -t & 0 & 0 & 0 & 0 & 0\\
0 & -t & 0 & -t & 0 & 0 & 0 & 0\\
-\Delta^* & 0 & -t & 0 & 0 & 0 & 0 & 0\\
0 & 0 & 0 & 0 & 0 & -t & 0 & -\Delta\\
0 & 0 & 0 & 0 & -t & 0 & -t & 0\\
0 & 0 & 0 & 0 & 0 & -t & 0 & -\Delta\\
0 & 0 & 0 & 0 & -\Delta^* & 0 & -\Delta^* & 0
\end{pmatrix}.
\end{equation}
Notice that only states with the same parity in the number of particles are coupled to each other. In this sense, there are two independent subspaces in the total Hilbert space, one consisting of states with an odd number of particles and another with an even number of particles, which we will call "odd" and "even" subspaces, respectively. Furthermore, if we take a closer look at the Hamiltonian, we can see that it is block diagonal with an identical periodic tight-binding Hamiltonian on each block. Due to these properties, we can establish a correspondence between the three-site Kitaev chain and two independent tight-binding rings with four sites each. The Hamiltonian of the system becomes

\begin{eqnarray}
&H&=H_{\text{even}}+H_\text{odd},\nonumber
\\
&H_{\text{even}}& = -\Delta\ket{1}\bra{2}-t\ket{2}\bra{3}-t\ket{3}\bra{4}-\Delta^*\ket{4}\bra{1}+H.c.,\nonumber
\\
&H_\text{odd}& = -t\ket{5}\bra{6}-t\ket{6}\bra{7}-\Delta\ket{7}\bra{8}-\Delta^*\ket{8}\bra{5}+H.c.\nonumber
\\
\end{eqnarray}
We have renamed the vectors in basis $B$ to simplify the writing and so, hereafter, $B=\{\ket{1},\ket{2},\ket{3},\ket{4},\ket{5},\ket{6},\ket{7},\ket{8}\}$. A pictorial representation of the system can be seen in Fig. \ref{TB_4}.

Having made the correspondence between both problems, we now find out the specific the correspondences between the states of the tight-binding chain and the states of the Kitaev chain. We will study only the case where $t=|\Delta|$, so $\Delta=te^{i\phi}$, where $\phi$ is the superconducting phase. Let us start by finding the eigenstates of the tight-binding problem. 

\paragraph{Eigenstates of the tight-binding rings}
The eigenstates of an Hamiltonian with translational invariance are obtained using Bloch's theorem, as it was already discussed in Chapter 2. The Hamiltonian of the tight-binding rings will have translational invariance if we perform the following gauge transformations

\begin{equation}
\begin{split}
e^{i\phi}\ket{1}\rightarrow\ket{\tilde{1}}
,\qquad
e^{-i\phi}\ket{8}\rightarrow\ket{\tilde{8}}.
\end{split}
\end{equation}
The new basis becomes $\tilde B=\{\ket{\tilde{1}},\ket{2},\ket{3},\ket{4},\ket{5},\ket{6},\ket{7},\ket{\tilde{8}}\}$ and we define \hspace{0.2cm}$\tilde B_{even}=\{\ket{\tilde{1}},\ket{2},\ket{3},\ket{4}\}$ and $\tilde B_{odd}=\{\ket{5},\ket{6},\ket{7},\ket{\tilde{8}}\}$. Using Bloch's theorem, the eigenstates of ring "even" and ring "odd" are, respectively,

\begin{equation}
\ket{k_{even}}=\frac{1}{2}\sum_{j_1}e^{ik_{even}j_1}\ket{j_1}
,\qquad
\ket{k_{odd}}=\frac{1}{2}\sum_{j_2}e^{ik_{odd}j_2}\ket{j_2},
\end{equation}
where $j_1$ and $j_2$ run through all the states in basis $\tilde B_{even}$ and $\tilde B_{odd}$, respectively, and $k_{even}$ and $k_{odd}$ are contained in the first Brillouin zone.
\paragraph{Eigenstates of the Kitaev chain}
Now let us find the states of the Kitaev chain and express them in basis $B$. As it was discussed above, the state of the Kitaev chain is obtained by specifying the action of the quasiparticle creation operators in the vacuum of the system.  Knowing this, let us start by finding the vacuum state written in basis $B$ and build the other states of the system from that. Hereafter, we will drop the phase in $\Delta$, so $\phi=0$, because it does not make a difference in the solutions of the problem. In fact, we do not need to worry about it because it can be absorbed in the definition of the Majorana operators,
\begin{equation}
\gamma_{j,1}=e^{-i\frac{\phi}{2}}c^\dagger_j+e^{i\frac{\phi}{2}}c_j,
\qquad \gamma_{j,2}=ie^{-i\frac{\phi}{2}}c^\dagger_j-ie^{i\frac{\phi}{2}}c_j,
\end{equation} 
yielding the same solution as in \ref{eq_5}. 
\paragraph{Finding the Vacuum State}
The vacuum state $\ket{\emptyset_{vac}}$ obeys the following set of equations

\begin{equation}
\tilde{c}_1\ket{\emptyset_{vac}}=0,
\qquad
\tilde{c}_2\ket{\emptyset_{vac}}=0,
\qquad...
\qquad
\tilde{c}_{N-1}\ket{\emptyset_{vac}}=0,
\qquad
\tilde{c}_M\ket{\emptyset_{vac}}=0.
\end{equation}
To find the solution written in terms of the vectors of basis $B$, we start by expressing the $\tilde{c}_j$ operators in terms of the original fermionic operators ${c}_j$ and ${c}_j^\dagger$:

\begin{equation}
\tilde{c}_j=\frac{1}{2}(\gamma_{j+1,1}+i\gamma_{j,2})=\frac{1}{2}(c_{j+1}^\dagger+c_{j+1}-c_j^\dagger+c_j);
\quad
\tilde{c}_M=\frac{i}{2}(c_{N}^\dagger-c_N+c_1^\dagger+c_1).
\end{equation}
Since basis $B$ is a complete set of orthonormal states, we are able to express the vacuum as

\begin{equation}
\ket{\emptyset_{vac}}=\sum_j \braket{j|\emptyset_{vac}}\ket{j},
\end{equation}
where $j$ runs through all the basis vectors and $\braket{j|\emptyset_{vac}}$ are coefficients. In matrix form and written in base $B$, the vacuum is

\begin{equation}
	\ket{\emptyset_{vac}}=\begin{pmatrix}
	a\\b\\c\\d\\e\\f\\g\\h\end{pmatrix},\qquad
	\braket{\emptyset_{vac}|\emptyset_{vac}}=1
\end{equation}
where the letters are coefficients. Because we know the action of the original fermionic operators on basis $B$ we can solve the following equations for the case of the Kitaev chain of 3 sites

\begin{equation}
\begin{split}
\tilde{c}_1\ket{\emptyset_{vac}}=\frac{1}{2}(c_{2}^\dagger+c_{2}-c_1^\dagger+c_1)\left(\sum_j \braket{\emptyset_{vac}|j}\ket{j}\right)=0,\\
\tilde{c}_2\ket{\emptyset_{vac}}=\frac{1}{2}(c_{3}^\dagger+c_{3}-c_2^\dagger+c_2)\left(\sum_j \braket{\emptyset_{vac}|j}\ket{j}\right)=0,\\
\tilde{c}_M\ket{\emptyset_{vac}}=\frac{1}{2}(c_{3}^\dagger-c_{3}+c_1^\dagger+c_1)\left(\sum_j \braket{\emptyset_{vac}|j}\ket{j}\right)=0.
\end{split}
\label{eq_0}
\end{equation}
The solution for this set of equations is\footnote{We chose a descending order of operation of the fermionic operators, so that, for example, we act with ${c}_2^\dagger$ before we act with ${c}_1^\dagger$}

\begin{equation}
\left\{\begin{matrix}
a=-b\\
c=-d\\
e=-f\\
g=-h
\end{matrix}\right.
,\qquad
\left\{\begin{matrix}
a=-d\\
b=-c\\
e=-h\\
f=-g
\end{matrix}\right.
,\qquad
\left\{\begin{matrix}
a=-c\\
b=-d\\
e=g\\
f=h
\end{matrix}\right.
.
\end{equation}
The only possible normalized solution, apart from a global phase factor, is given by

\begin{equation}
\ket{\emptyset_{vac}}=\frac{1}{2}\begin{pmatrix}
0\\0\\0\\0\\+1\\-1\\+1\\-1\end{pmatrix},
\end{equation}

\paragraph{Eigenstates of the system}
Since we are now working on the eigenbasis of the Hamiltonian, we can construct the eigenstates of the system by acting in all possible configurations with the creation operators on the vacuum state. The eigenstates will be,

\begin{equation}
\ket{\emptyset_{vac}},
\qquad \tilde{c}_M^\dagger\ket{\emptyset_{vac}},
\qquad \tilde{c}_1^\dagger\ket{\emptyset_{vac}},
\qquad\tilde{c}_M^\dagger\tilde{c}_1^\dagger\ket{\emptyset_{vac}}
\qquad...
\qquad\tilde{c}_M^\dagger\tilde{c}_2^\dagger\tilde{c}_1^\dagger\ket{\emptyset_{vac}}.
\end{equation}
Written in matrix form, the states are,

\small
\begin{multline}
\ket{\emptyset_{vac}}=\frac{1}{2}\begin{pmatrix}
0\\0\\0\\0\\+1\\-1\\+1\\-1\end{pmatrix},\hspace{0.12cm}	\tilde{c}_M^\dagger\ket{\emptyset_{vac}}=\frac{i}{2}\begin{pmatrix}
-1\\+1\\-1\\+1\\0\\0\\0\\0\end{pmatrix},\hspace{0.12cm}
\tilde{c}_1^\dagger\ket{\emptyset_{vac}}=\frac{1}{2}\begin{pmatrix}
-1\\-1\\+1\\+1\\0\\0\\0\\0\end{pmatrix},\hspace{0.12cm}
\tilde{c}_1^\dagger\tilde{c}_M^\dagger\ket{\emptyset_{vac}}=\frac{i}{2}\begin{pmatrix}
0\\0\\0\\0\\-1\\-1\\+1\\+1\end{pmatrix}\\
\tilde{c}_2^\dagger\ket{\emptyset_{vac}}=\frac{1}{2}\begin{pmatrix},
+1\\-1\\-1\\+1\\0\\0\\0\\0\end{pmatrix},\hspace{0.12cm}	\tilde{c}_2^\dagger\tilde{c}_M^\dagger\ket{\emptyset_{vac}}=\frac{i}{2}\begin{pmatrix}
0\\0\\0\\0\\+1\\-1\\-1\\+1\end{pmatrix},\hspace{0.12cm}
\tilde{c}_1^\dagger\tilde{c}_2^\dagger\ket{\emptyset_{vac}}=\frac{1}{2}\begin{pmatrix}
0\\0\\0\\0\\+1\\+1\\+1\\+1\end{pmatrix},\hspace{0.12cm}
\tilde{c}_1^\dagger\tilde{c}_2^\dagger\tilde{c}_M^\dagger\ket{\emptyset_{vac}}=\frac{i}{2}\begin{pmatrix}
-1\\-1\\-1\\-1\\0\\0\\0\\0\end{pmatrix}
\end{multline}
\normalsize

Notice that every energy is at least doubly degenerate, as expected, so any linear combination of eigenstates with the same energy will also be an eigenstate. Knowing this, we can explicitly establish the correspondence between the states above and the states of the tight binding rings constructed for the "even" and "odd" subspaces.

\paragraph{State-site correspondence}
The states of the tight binding rings are the following\footnote{Notice that we dropped the phase factor in $\Delta$, so $\ket{1}=\ket{\tilde{1}}$},

\begin{eqnarray}
\ket{k_{even}=0}&=&\frac{1}{2}(\ket{1}+\ket{2}+\ket{3}+\ket{4}),\qquad
\ket{k_{even}=\frac{\pi}{2}}=\frac{1}{2}(i\ket{1}-\ket{2}-i\ket{3}+\ket{4}),
\nonumber
\\
\ket{k_{even}=\pi}&=&\frac{1}{2}(-\ket{1}+\ket{2}-\ket{3}+\ket{4}),\qquad
\ket{k_{even}=-\frac{\pi}{2}}=\frac{1}{2}(-i\ket{1}-\ket{2}+i\ket{3}+\ket{4}),
\nonumber
\\
\ket{k_{odd}=0}&=&\frac{1}{2}(\ket{5}+\ket{6}+\ket{7}+\ket{8}),\qquad
\ket{k_{odd}=\frac{\pi}{2}}=\frac{1}{2}(i\ket{5}-\ket{6}-i\ket{7}+\ket{8}),
\nonumber
\\
\ket{k_{odd}=\pi}&=&\frac{1}{2}(-\ket{5}+\ket{6}-\ket{7}+\ket{8}),\qquad
\ket{k_{odd}=-\frac{\pi}{2}}=\frac{1}{2}(-i\ket{5}-\ket{6}+i\ket{7}+\ket{8}).\nonumber\\
\end{eqnarray}
Just by looking at the states above one can directly establish the following correspondences,

\small
\begin{eqnarray}
-i\ket{k_{even}=0}&\rightarrow&\tilde{c}_1^\dagger\tilde{c}_2^\dagger\tilde{c}_M^\dagger\ket{\emptyset_{vac}},\qquad
\ket{k_{odd}=0}\rightarrow\tilde{c}_1^\dagger\tilde{c}_2^\dagger\ket{\emptyset_{vac}},\\
\nonumber
-i\ket{k_{even}=\pi}&\rightarrow&\tilde{c}_M^\dagger\ket{\emptyset_{vac}},\qquad\
-\ket{k_{odd}=\pi}\rightarrow\ket{\emptyset_{vac}},
\end{eqnarray}
\normalsize
and, since any linear combination of eigenstates  with the same energy are also eigenstates, one possible case of the remaining correspondences are given by
	 
\begin{eqnarray}
\frac{1+i}{2}&\ket{k_{even}=\frac{\pi}{2}}&+\frac{1-i}{2}\ket{k_{even}=-\frac{\pi}{2}}\rightarrow\tilde{c}_1^\dagger\ket{\emptyset_{vac}},
\\
\frac{1-i}{2}&\ket{k_{even}=\frac{\pi}{2}}&+\frac{1+i}{2}\ket{k_{even}=-\frac{\pi}{2}}\rightarrow\tilde{c}_2^\dagger\ket{\emptyset_{vac}},
\\
\frac{-1+i}{2}&\ket{k_{odd}=\frac{\pi}{2}}&+\frac{1+i}{2}\ket{k_{odd}=-\frac{\pi}{2}}\rightarrow\tilde{c}_1^\dagger\tilde{c_M}^\dagger\ket{\emptyset_{vac}},
\\
\frac{1+i}{2}&\ket{k_{odd}=\frac{\pi}{2}}&+\frac{-1+i}{2}\ket{k_{odd}=-\frac{\pi}{2}}\rightarrow\tilde{c}_2^\dagger\tilde{c_M}^\dagger\ket{\emptyset_{vac}}.
\end{eqnarray}
In general, the eigenstates of the Kitaev chain correspond to linear combinations of the eigenstates of the tight-binding ring with the same energy.

\section{Generalization of the Kitaev $\Leftrightarrow$ Tight-binding correspondence}

In this section we will generalize the concept of establishing a correspondence between the Kitaev chain problem and a tight-binding problem for any number of sites $N$. This will be done by providing a rule of construction of the tight-binding lattices and a rule of attributing the states of the original basis of the Kitaev chain to the sites of the tight binding lattices.

The general tight-binding problem that corresponds to an $N$ site Kitaev chain consists of two independent and identical lattices that have the form of $2^{N-1}$ side regular polygons. The sites of one of these lattices are associated only with states of even parity and the sites of the other are associated only with states of odd parity and, since they are identical and share the same 
\begin{figure}
	\centering
	\includegraphics[scale=0.35]{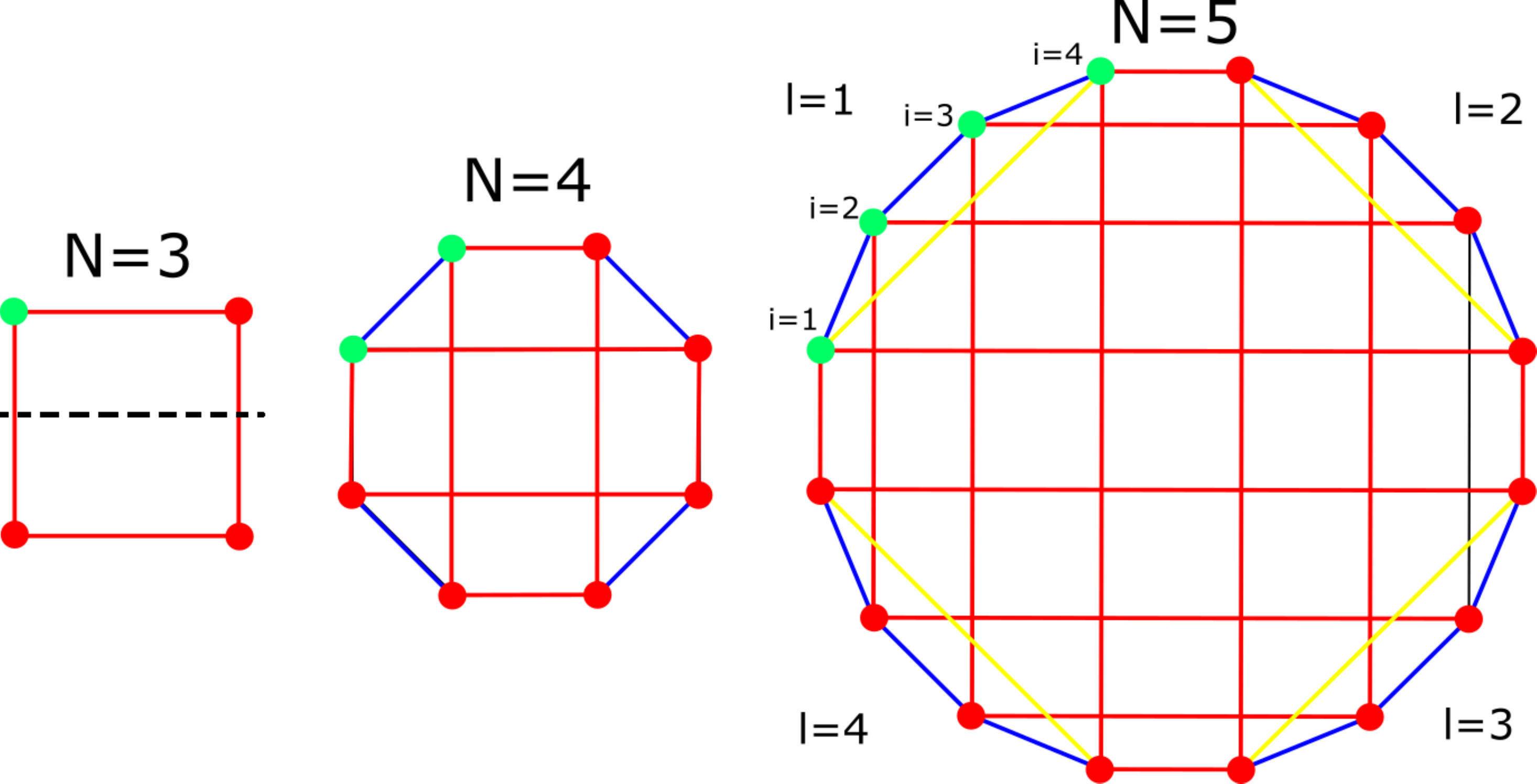}
	\caption{Tight-binding lattices correspondent to the $N=3$, $N=4$ and $N=5$ sites Kitaev problem, respectively. The sites of the unit cell are represented in green. The other colors represent the different kinds of hoppings. In the $N=5$ lattice we also represent the indexes of the unit cells and of the sites of the unit cell in equation \eqref{hamil}, to facilitate the reading. Notice that by cutting the chain with $N$ sites in half horizontally (as exemplified for $N=3$ by cutting along the dashed line) one obtains the unit cell of the $N+1$ chain. The sites with $i=1$ are the first sites in unit cell $l$.}
	\label{figura1}
\end{figure}
rules of construction, we will hereafter be describing only the "even" lattice, keeping in mind that an identical lattice describes the "odd" subspace. The lattices are always constituted by four unit cells of $2^{N-3}$ sites each.

One interesting property of these lattices, and also one that will allow us to describe them for any $N$, is that the unit cell of the lattice correspondent to the $N+1$ sites Kitaev problem is obtained by cutting in half the lattice of the $N$ sites Kitaev problem, as can be checked by looking at Fig. \ref{figura1}. We distinguish between three types of connections in these lattices. The first type connects nearest-neighboring sites in the same unit cell, the second connects sites in the same unit cell that are not nearest-neighbors and the third connects different unit cells, each of these connections being represented in a different color in Fig. \ref{figura1}.

\begin{figure}
	\centering
	\includegraphics[scale=0.5]{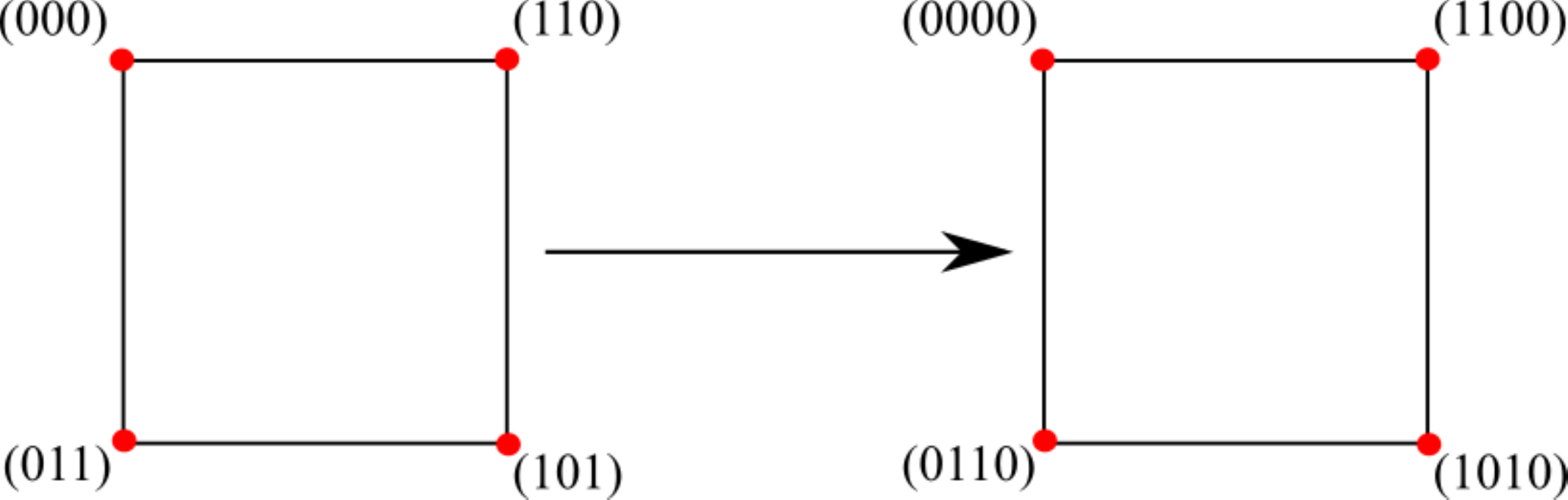}
	\caption{The left square is the tight-binding lattice associated with the "even" subspace for $N=3$ with the state-site correspondences already made. The right square represents the first step in transitioning from the $N=3$ to the $N=4$ chain, where a 0 is added at the end of the label of each state.}
	\label{3sitios}
\end{figure}

\begin{figure}
	\centering
	\includegraphics[scale=0.4]{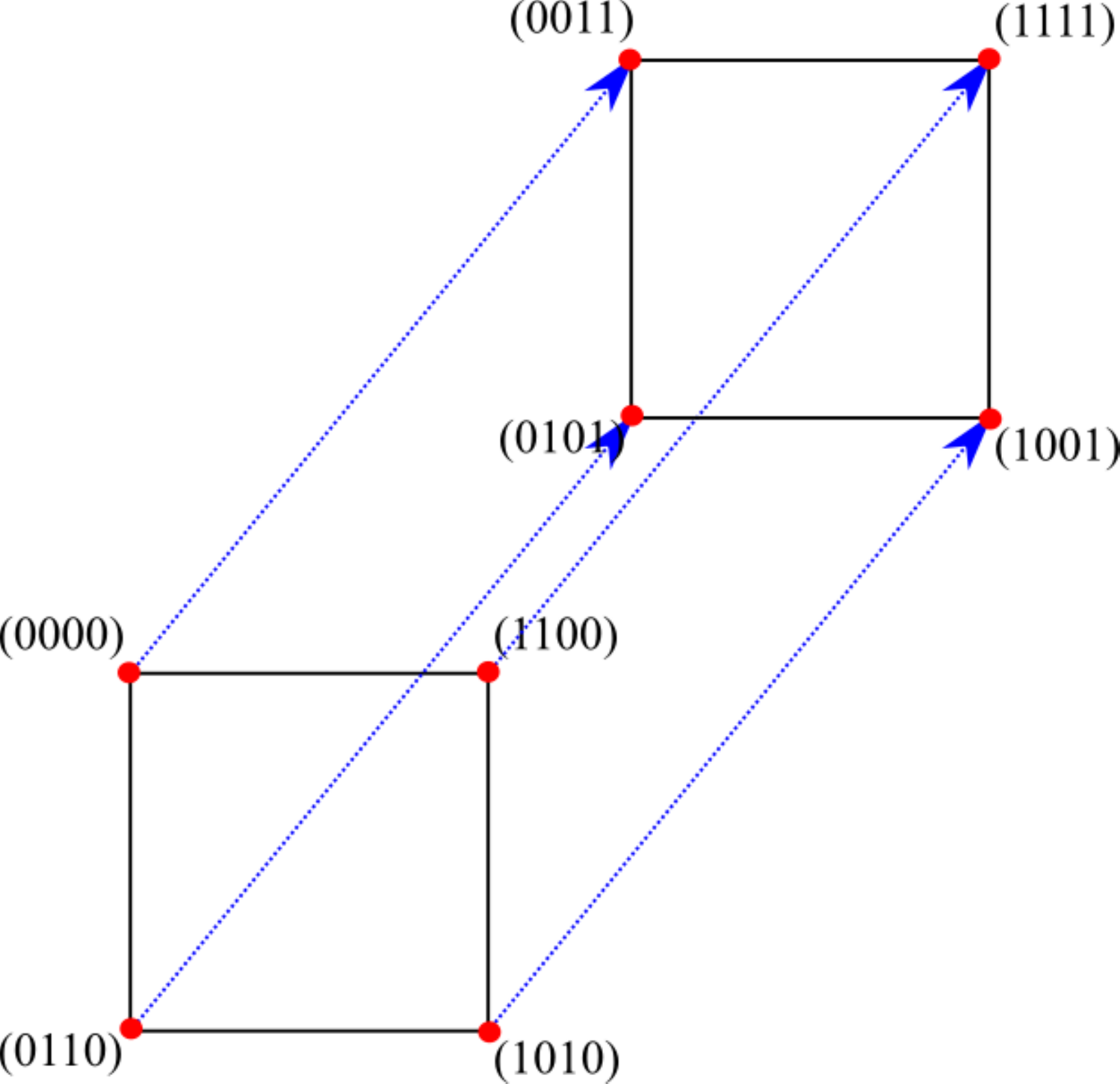}
	\caption{The shape found in Fig. \ref{3sitios} is pulled outwards and creates this three-dimensional shape. The missing connections between sites for the $N=4$ case are represented here as blue arrows.}
	\label{azul}
\end{figure}

Before we present the generalized Hamiltonian that describes the tight-binding lattices, let us first provide the rule that allows one to attribute the states of the Kitaev chain to the sites of these lattices. This rule allows us to discover the state-site correspondences for the $N+1$ case knowing them for the $N$ case. We start by considering one of the lattices for $N=3$ with the state-site correspondences already made in section \ref{threesite}, as shown in the left square of Fig. \ref{3sitios}. If we now add one zero to the end of the label of every state (see right square of Fig. \ref{3sitios}) the state-site correspondences obtained are still valid, only now they are linking states of the $N=4$ Kitaev chain. When we add one site to the Kitaev chain each state of the basis couples with one more state, so there is still one connection missing for each site. However, there is only one possible state that can couple to each of the states of the right square in Fig. \ref{3sitios}, since two out of three connections are already established, and these missing states will couple with each other in a similar fashion to the ones shown in the figure. So, we can imagine that we duplicate the square and then pull it outwards , so that each state couples with the one directly above it, as exemplified in Fig. \ref{azul}. The label of the duplicated states  follows the rule

\begin{equation}
\text{The state } (...x0), \text{ links to } (...y1)\qquad
\left\{\begin{matrix}
\text{if } x=0\implies y=1\\
\text{if } x=1\implies y=0
\end{matrix}\right.
.
\label{rule}
\end{equation}
This is exemplified in Fig.\ref{azul}. The only thing left to obtain the $N=4$ tight-binding lattice of Fig. \ref{figura1} is to project this shape into a two dimensional plane. 

\begin{figure}
	\centering
	\includegraphics[scale=0.4]{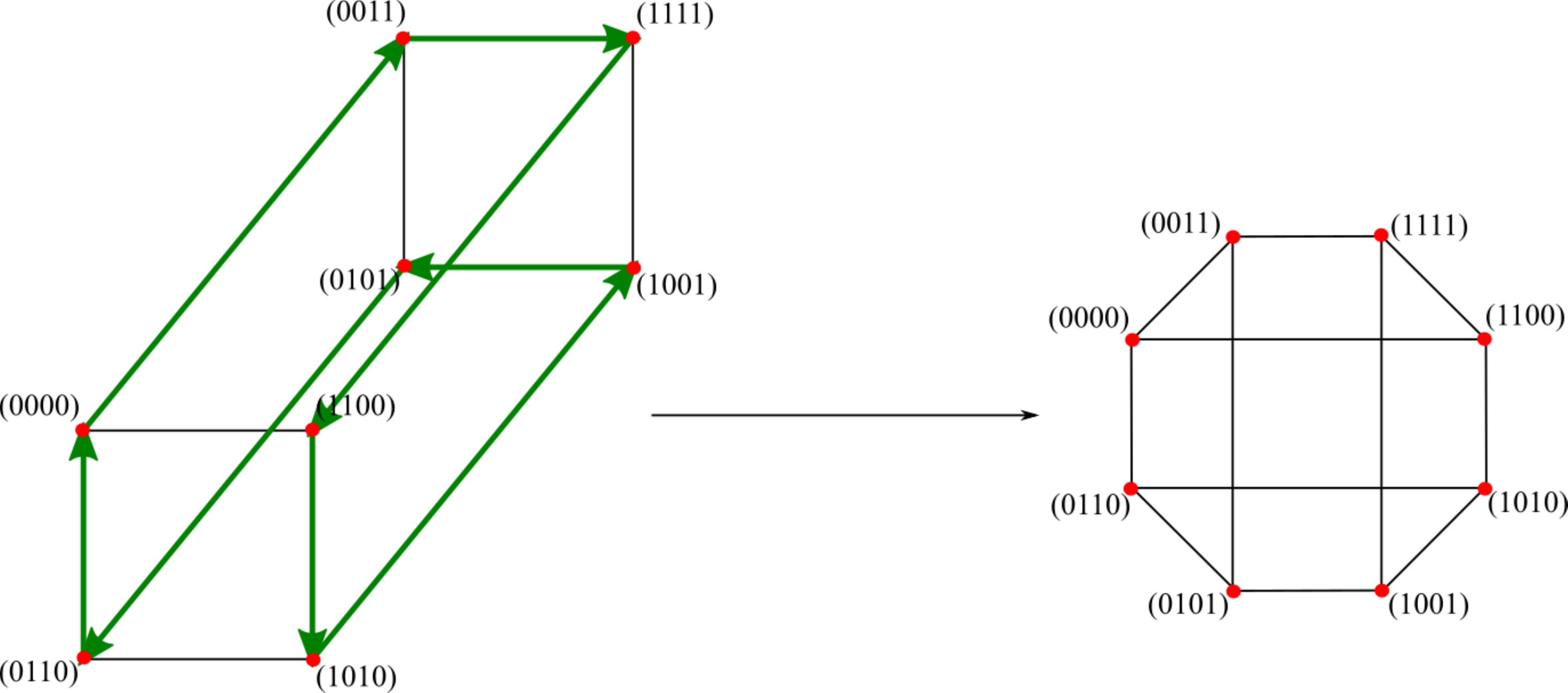}
	\caption{The path taken along the shape is shown by the arrows. The resulting tight-binding lattice is shown at the right.}
	\label{verde}
\end{figure}

We do so by following a simple rule, which consists of taking one of  the longest paths that starts in the first site of any unit cell and ends in the same site along the 3-dimensional shape created by pulling the square outwards and, by each site we pass through we attribute it in clockwise order to the vertexes of a $2^{N-1}$ side polygon, which is precisely the shape of the tight-binding lattice of the $N+1$ Kitaev chain. Let us follow this rule for the case of $N=4$. 
We take the site correspondent to the vacuum state $(0000)$\footnote{Here, the notation  $(0000)$ stands for $\ket{0000}$.}, as the site where we start taking the path from. Now if we repeat the pattern up-clockwise turn-down-clockwise turn, along the shape at the left of Fig. \ref{verde} and attribute each site we pass through to the polygon in clockwise order we get the state-site correspondences for $N=4$, as shown at the right of Fig. \ref{verde}. To find the shape of the $N=5$ tight-binding chain, show in Fig. \ref{figura1}, one has to follow the same pattern, that is, duplicate the octagon and pull it outwards, labeling the states by the rule \eqref{rule} and then take the longest path starting at the first site of any unit cell along the duplicated octagon and, when projected into a plane, we find the $N=5$ shape of the tight-binding lattice. The process is repeated to find the shapes for higher $N$.

Given this repeating pattern, it is possible to present a generalized Hamiltonian of the tight-binding lattices for any number of sites $N$ of the Kitaev chain,

\begin{multline}
H=-t\sum_{l=1}^{4}
\Biggl(
\sum_{i=1}^{N_{uc}-1}(c^\dagger_{l,i}c_{l,i+1})+\sum_{i=1}^{N_{uc}}(c^\dagger_{l,i}c_{l+1,N_{uc}+1-i})+\\
\sum_{j=1}^{N_d-2}\sum_{u=1}^{\frac{1}{2}d_j-1}\sum_{s=0}^{(d_{N_d+1-j})-1}(c^\dagger_{l,u+sd_j}c_{l,d_j(1+s)+1-u})
+H.c.\Biggr),
\label{hamil}
\end{multline}
with $N_{uc}$ the number of sites in the unit cell, $d_j$ the \textit{j}th divisor of $N_{uc}$, $N_d$ is the number of divisors\footnote{The first divisor of $N_{uc}$ is $N_{uc}$ itself and 1 is the last divisor. For example, for $N_{uc}=4$ one has $d_1=4$, $d_2=2$, $d_3=1$, so $N_d=3$} and $c^\dagger_{l,i}$ is the creation operator associated to site $i$ of unit cell $l$. This expression is only valid for $N>4$ but the Hamiltonian's for $N\leq4$ are easy to find. The first sum in \eqref{hamil} runs through all the unit cells. The first term in the parenthesis expresses the hopping between nearest-neighboring sites in the same unit cell whereas the second term in the parenthesis expresses the hopping between sites of different adjacent unit cells. Lastly, the third term in the parenthesis expresses the hopping between sites in the same unit cell that are not nearest-neighboring sites.

\newpage
\section{Band structure of the tight-binding chain}
Finally, in this section, we calculate the band structure of the tight-binding lattice by expressing the Hamiltonian in $k$-space for any number of sites $N$ of the Kitaev chain. Let us start by considering the Hamiltonian of the $N=4$ tight-binding chain,

\begin{equation}
H=-t\sum_{j=1}^{4} \left(c_{1,j}^\dagger c_{2,j+1}+c_{1,j}^\dagger c_{2,j}+c_{2,j}^\dagger c_{1,j+1}+H.c.\right),
\label{hamil_2}
\end{equation}
where $c^\dagger_{i,j}$ is the creation operator associated to site $i$ of unit cell $j$. Using Bloch's theorem, we have

\begin{equation}
c_{1,j}^\dagger=\frac{1}{\sqrt{N}}\sum_k e^{-ikj}c^\dagger_{1,k},\qquad
c_{2,j}^\dagger=\frac{1}{\sqrt{N}}\sum_k e^{-ikj}c^\dagger_{2,k}.
\end{equation}
Substituting these relations in the Hamiltonian of \eqref{hamil_2} we get,

\begin{equation}
H=-t\sum_k (2\cos{k}+1)c^\dagger_{1,k}c_{2,k}+(2\cos{k}+1)c^\dagger_{2,k}c_{1,k}.
\end{equation}
Written in matrix representation the Hamiltonian becomes,

\begin{equation}
H=-t\sum_{k>0}\begin{pmatrix}c^\dagger_{1,k}&c^\dagger_{2,k}\end{pmatrix}\begin{pmatrix}0&\alpha_k\\
\alpha_k& 0\end{pmatrix}\begin{pmatrix}c_{1,k}\\
c_{2,k}\end{pmatrix},
\end{equation}
where $\alpha_k=1+2\cos{k}$ and $k=\frac{2\pi n}{4}$, where $n$ is the same as in \eqref{eq_fixe}. Analogously, we can find the Hamiltonian for any $N$. For instance, the results for $N=5$ and $N=6$ are

\begin{figure}[H]
	\centering
	\includegraphics[scale=0.50]{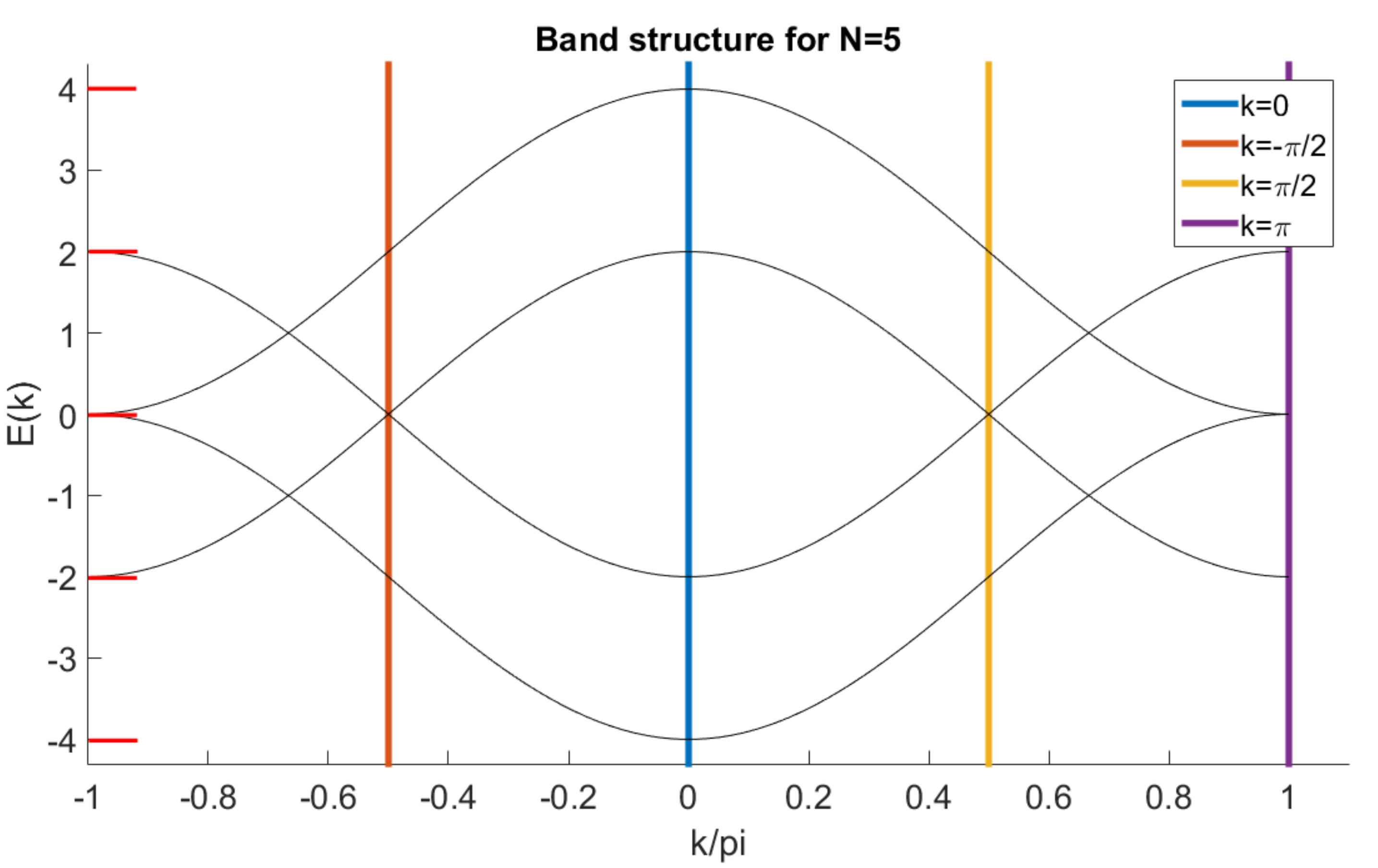}
	\caption{Band structure of the "even" tight-binding lattice correspondent to the $N=5$ Kitaev chain. The energy levels for the allowed $k$ values are given by the intersection of the vertical lines with the graph. The red lines mark the energy levels for $t=\Delta$.}
	\label{Band_TB}
\end{figure}

\small
\begin{equation}
H=-t\begin{pmatrix}0&1&0&\alpha_k\\
1&0&\alpha_k&0\\
0&\alpha_k&0&1\\
\alpha_k&0&1&0\\
\end{pmatrix}
,\text{for N=5},
\hspace{0.1cm}
H=-t\begin{pmatrix}
0&1&0&1&0&0&0&\alpha_k\\
1&0&1&0&0&0&\alpha_k&0\\
0&1&0&1&0&\alpha_k&0&0\\
1&0&1&0&\alpha_k&0&0&0\\
0&0&0&\alpha_k&0&1&0&1\\
0&0&\alpha_k&0&1&0&1&0\\
0&\alpha_k&0&0&0&1&0&1\\
\alpha_k&0&0&0&1&0&1&0\\
\end{pmatrix}
,\text{for N=6}.
\end{equation}
\normalsize
Notice that the matrix representation of the Hamiltonian is following a pattern. The elements of the anti-diagonal are always $\alpha_k$ and the Hamiltonian for $N+1$ copies the shape of the Hamiltonian for $N$ in the diagonal blocks (substituting $\alpha_k$ for 1). This repeating pattern allows us to write a generalized expression for the Hamiltonian in $k$-space for any given $N$,

\begin{equation}
H=-t\left( \alpha_k(\prod_\otimes^{N-3}\sigma_x)+\sum_{i=1}^{N-4}\left((\prod_\otimes^{i}\sigma_x)\otimes(\prod_\otimes^{N-3-i} \sigma_0)\right)\right) 
\end{equation}
where $\sigma_0$ is the $2\times2$ identity matrix, $\sigma_x$ is the first Pauli matrix and the notation $\prod_\otimes^{N}\sigma_x$ stands for $\sigma_x\otimes\sigma_x...\otimes\sigma_x\text{ N times}$. Regardless of the $N$ considered, the correspondent tight-binding lattices will always be periodic with four unit cells (see Fig. \ref{figura1}). Therefore, the allowed $k$-states in the band structure for every $N$ are always $k=0,\pm \frac{\pi}{2},\pi$. In Fig. \ref{Band_TB}, we plot the energy spectrum of one of the tight-binding lattices correspondent to the $N=5$ Kitaev chain. The energies at the allowed $k$-states are consistent with those found in Fig. \ref{figura3} for $t=\Delta$, as expected.

\chapter{Conclusion}
In this project we studied the Kitaev chain, a simple toy model that exhibits unpaired Majorana fermions.
We started by taking Kitaev's approach to find the exact solution of the problem for the case $t=|\Delta|$, where we found the presence of non-local zero energy states. We then studied the energy spectrum of the bulk of the chain, drawing conclusions on the different topological phases of the system.
We performed numerical calculations to explore the energy levels of the chain for different parameters $t$, $\Delta$ and $\mu$, and we concluded that there are three regimes where we know the exact solution of the problem. 

We further explored the regime $t=|\Delta|$, making a correspondence between the Kitaev chain for $N=3$ and a tight-binding problem consisting of two identical and independent rings with four sites each. Building on that, we generalized the correspondence between the Kitaev chain and a tight-binding problem for an arbitrary number of sites, providing the rules of construction of the tight-binding lattices. Finally, we studied the energy spectrum of the tight-binding lattices and presented a general expression for the matrix representation of the Hamiltonian in $k$-space.

The correspondence of the Kitaev chain with two tight-binding rings will allow, in principle, to understand more easily the modifications of the behavior of the Kitaev chain due to possible perturbations. In particular, it would be interesting to find a perturbation of the Kitaev chain that maps onto a magnetic flux in the tight-binding rings.

\bibliographystyle{unsrt}

\bibliography{bibliography}

\end{document}